\documentclass[5p]{elsarticle}

\usepackage{lineno,hyperref}
\modulolinenumbers[1]

\usepackage{amsmath,amsfonts,amssymb}
\usepackage{graphicx}
\usepackage{aas_macros}
\usepackage{multirow}
\usepackage{subfig}

\begin{document}

 \begin{frontmatter}
 
 \title{The IXPE Instrument Calibration Equipment}

\author[1]{Fabio Muleri\corref{corr}}
\author[1]{Raffaele Piazzolla}
\author[1]{Alessandro Di Marco}
\author[1]{Sergio Fabiani}
\author[1]{Fabio La Monaca}
\author[1]{Carlo Lefevre}
\author[1]{Alfredo Morbidini}
\author[1,2,3]{John Rankin}
\author[1]{Paolo Soffitta}
\author[1]{Antonino Tobia}
\author[1]{Fei Xie}
\author[1]{Fabrizio Amici}
\author[4]{Primo Attinà}
\author[5]{Matteo Bachetti}
\author[1]{Daniele Brienza}
\author[6]{Mauro Centrone}
\author[1]{Enrico Costa}
\author[1]{Ettore Del Monte}
\author[1]{Sergio Di Cosimo}
\author[1]{Giuseppe Di Persio}
\author[1]{Yuri Evangelista}
\author[1,2,3]{Riccardo Ferrazzoli}
\author[1]{Pasqualino Loffredo}
\author[6]{Matteo Perri}
\author[5]{Maura Pilia}
\author[7]{Simonetta Puccetti}
\author[1,2,3]{Ajay Ratheesh}
\author[1]{Alda Rubini}
\author[1]{Francesco Santoli}
\author[1]{Emanuele Scalise}
\author[5]{Alessio Trois}
\cortext[corr]{Corresponding author}
\address[1]{Istituto di Astrofisica e Planetologia Spaziali di Roma, Via Fosso del Cavaliere 100, I-00133 Roma, Italy}
\address[2]{Università di Roma Sapienza, Dipartimento di Fisica, Piazzale Aldo Moro 2, 00185 Roma, Italy}
\address[3]{Università di Roma Tor Vergata, Dipartimento di Fisica, Via della Ricerca Scientifica, 1, 00133 Roma, Italy}
\address[4]{INAF/Osservatorio Astrofisico di Torino, Via Osservatorio 20, I-10025 Pino Torinese (TO), Italy}
\address[5]{INAF/Osservatorio Astronomico di Cagliari, Via della Scienza 5, I-09047 Selargius (CA), Italy}
\address[6]{INAF/Osservatorio Astronomico di Roma, Via Frascati 33, I-00040, Monte Porzio Catone (RM)}
\address[7]{Agenzia Spaziale Italiana, Via del Politecnico snc, I-00133 Roma, Italy}

 \begin{abstract}
   The Imaging X-ray Polarimetry Explorer is a mission dedicated to the measurement of X-ray polarization from tens of astrophysical sources belonging to different classes. Expected to be launched at the end of 2021, the payload comprises three mirrors and three focal plane imaging polarimeters, the latter being designed and built in Italy. While calibration is always an essential phase in the development of high-energy space missions, for IXPE it has been particularly extensive both to calibrate the response to polarization, which is peculiar to IXPE, and to achieve a statistical uncertainty below the expected sensitivity. In this paper we present the calibration equipment that was designed and built at INAF-IAPS in Rome, Italy, for the calibration of the polarization-sensitive focal plane detectors on-board IXPE. Equipment includes calibration sources, both polarized and unpolarized, stages to align and move the beam, test detectors and their mechanical assembly. While all these equipments were designed to fit the specific needs of the IXPE Instrument calibration, their versatility could also be used in the future for other projects.
  \end{abstract} 

\end{frontmatter}

\section{Introduction}\label{sec:Introduction}

The Imaging X-ray Polarimetry Explorer (IXPE)\cite{Weisskopf2016} is the next NASA SMall EXplorer (SMEX) mission, to be launched at the end of 2021 and built in collaboration with the Italian Space Agency (Agenzia Spaziale Italiana, or ASI). IXPE, lead by NASA Marshall Space Flight Center (MSFC) in Huntsville, AL, features three identical telescopes, each composed of a grazing incidence mirror and a focal plane detector sensitive to X-ray polarization. The mission will measure the polarization of tens of astrophysical sources, both galactic and extragalactic, in the energy range between 2 and 8~keV and with good spectral, imaging and time resolution.

\begin{figure}
\begin{center}
\begin{tabular}{c}  
\includegraphics[width=6cm]{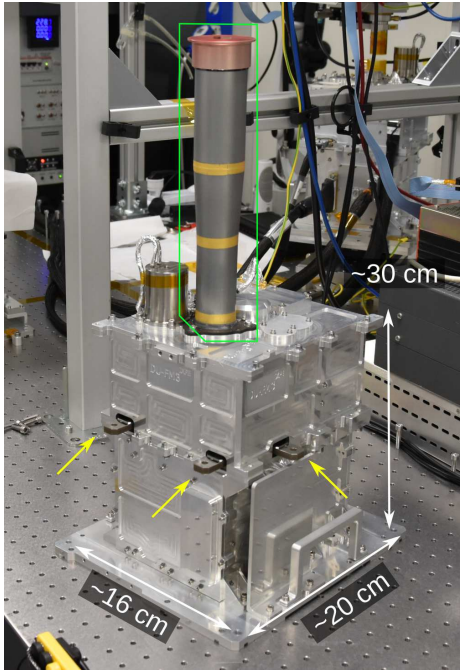}
\end{tabular}
\end{center}
\caption{One of the IXPE DU, manufactured in INFN. The straylight collimator (enclosed in the green box) is removed at the beginning of DU calibration to decrease the distance between the detector and the source (see Section~\ref{sec:Absorption}). Yellow arrows show the three mechanical references of the GPD for alignment.}
\label{fig:DU_reference} 
\end{figure}

The focal plane detectors on-board IXPE are based on the Gas Pixel Detector (GPD) design, which have been developed in Italy for nearly 20 years by a collaboration of Istituto Nazionale di Fisica Nucleare (INFN) and Istituto Nazionale di Astrofisica/Istituto di Astrofisica e Planetologia Spaziali (INAF-IAPS) in Rome \cite{Costa2001,Bellazzini2006,Bellazzini2007}. These detectors are the main Italian contribution to IXPE, which also includes the electronics to interface them to the spacecraft, the primary ground station and several contributions for the data processing pipeline, scientific analysis and data exploitation \cite{Soffitta2021}. IXPE polarimeters, named Detector Units (DUs, see Figure~\ref{fig:DU_reference}), were manufactured in INFN\cite{Baldini2021}, whereas the Detector Service Unit (DSU) which interfaces them to the spacecraft was built by OHB-Italia. A flight DSU and four DUs have been produced, three for flight plus one spare.

\begin{figure*}
\begin{center}
\begin{tabular}{c}
\includegraphics[width=17cm]{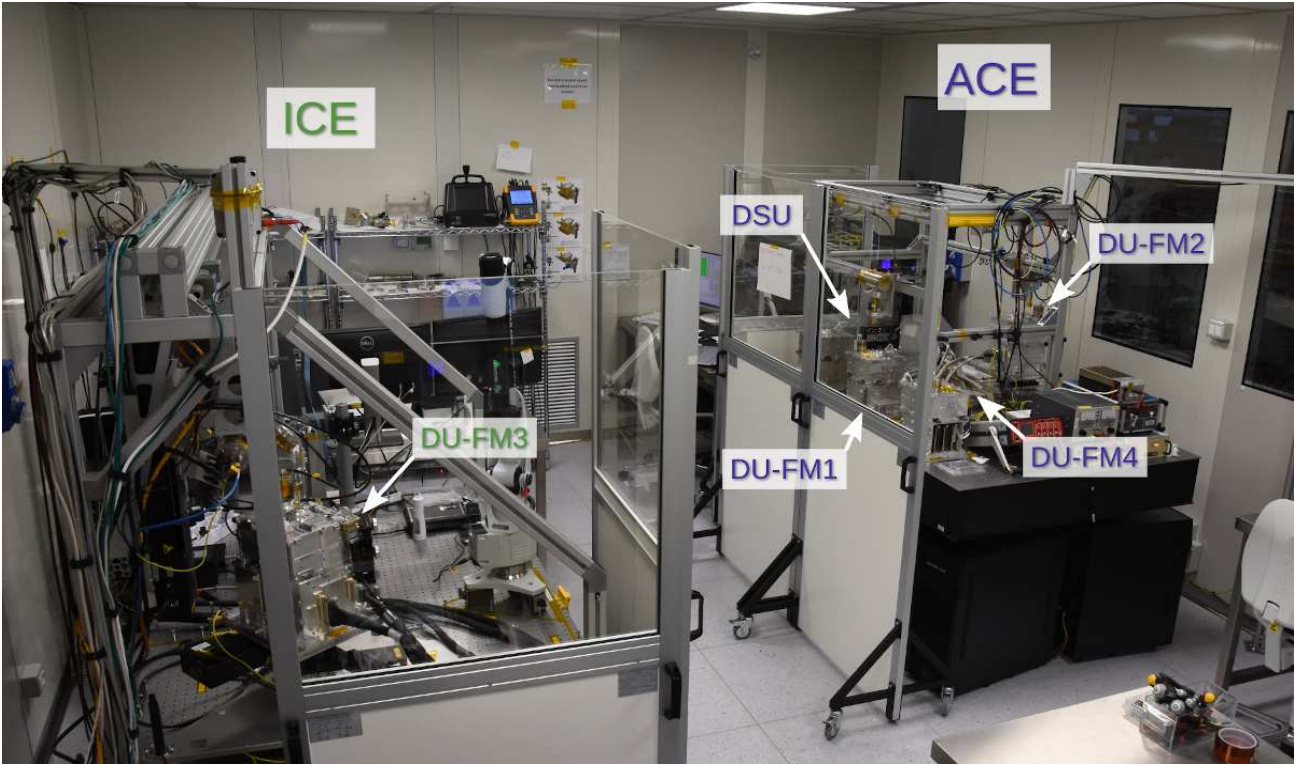}
\end{tabular}
\end{center}
\caption{Overview of the two calibration stations, named ICE and ACE, in the ISO~7 (class 10,000) IXPE clean room at INAF-IAPS during the contemporaneous calibration of the third DU flight model (DU-FM3) on the ICE and the Instrument testing with DSU-EM and DU-FM1, DU-FM2 and DU-FM4 on the ACE. Shields are used to limit the X-ray exposure of operators.}
\label{fig:ICE_ACE} 
\end{figure*}

The DUs and the DSU, comprehensively named the IXPE Instrument, were delivered to INAF-IAPS for extensive tests with X-rays before the integration on the spacecraft at Ball Aerospace, in Boulder, CO. Each DU, including both the flight and the spare units, went through a comprehensive calibration to finely characterize the response to both polarized and unpolarized radiation, and to measure their spectral, spatial and timing performance. The DUs were also integrated to the DSU and illuminated with X-ray sources to effectively test the operation of the whole IXPE Instrument in a configuration equivalent to the flight one\cite{Tobia2021}. All of these activities required equipment to generate X-rays with a polarization degree and angle precisely known. In this paper, we described the apparatus that we built for this purpose, which includes the X-ray calibration sources and the other electrical and mechanical items needed to operate them (see Figure~\ref{fig:ICE_ACE}).

It is worth noting here that the calibration of the focal plane detectors did not conclude the calibration of the IXPE observatory. Mirrors were separately calibrated at the Straylight Facility at the NASA-MSFC, and the spare DU and the spare Mirror were calibrated together again at the NASA-MSFC to confirm that calibration of the IXPE telescope could be extrapolated from the separate results obtained on the corresponding mirror and detector units. Moreover, each DU hosts a set of four calibration sources, which were used on-ground and which will be used in-orbit to monitor the response of the detector to both polarized and unpolarized sources\cite{Ferrazzoli2020}.

\section{IXPE Instrument calibration and testing}

Expectations from modeling of astrophysical sources which will be observed by IXPE require the instrumentation to resolve polarization of the order of 1\% or lower. This demands a characterization of the instrument response to better than such a value. Achieving such a goal is demanding statistically, as it requires the collections of tens of millions of photons for any energy of interest, and it requires the use of appropriate calibration sources. As a matter of fact, the only previous attempt to extensively calibrate an X-ray polarimeter was the Stellar X-Ray Polarimeter (SXRP)\cite{Tomsick1997}, which, however, had fairly different requirements with respect to IXPE because of the different sensitivity, energy range, lack of imaging capabilities and, ultimately, working principle of the detector. 

Calibration of the IXPE Instrument relied on custom calibration sources specifically designed and built in-house. The use of large synchrotron facilities, albeit possible at least for calibration with polarized radiation, was discouraged for the IXPE Instrument by a number of factors. On the one hand, the time allocated for each DU was 40 days, for a total of nominally 160 days for the three flight units and the spare one. This quite long time was necessary to collect a sufficient number of events and achieve the required sensitivity with a sustainable rate for the detector. In this context, the use of extremely bright sources was of little help, whereas planning the occupancy of a large facility for such a long time was a big challenge, also because the IXPE schedule could not be driven by calibration. On the other hand, an in-house system provided a great flexibility, as it was always available for the project, and facilitated the specialized measurements to be carry out on the unique detector on-board IXPE. All of this was essential to successfully carry out IXPE Instrument calibration in the allocated time interval.

Calibration of each DU consisted in a sequence of measurements, intended to characterize specific features of the instrument. These included:
\begin{itemize}
 \item the response of the instrument to unpolarized radiation. This needs to be subtracted from real observations as instrumental systematic effects, named collectively \emph{spurious modulation}, may mimic a real polarized signal. The requirement was to achieve a statistical uncertainty $\sigma$ on the knowledge of this component $<$0.1\%, repeating the measurement at six energies in the IXPE energy range. As $\sigma\approx\sqrt{2/N}$ \cite{Kislat2015}, the number of required counts is $N>2\cdot10^6$, which makes this kind of calibration the most time-consuming;
 \item the amplitude of the response to completely polarized X-rays, that is, the \emph{modulation factor}. In this case, measurements were carried out at 7~different energies as this quantity rapidly changes with energy;
 \item the absolute quantum efficiency of the detector with $<$5\% uncertainty. This requires a relatively modest number of counts, but a fine control of the systematic effect of the measurement, e.g., the knowledge of the detector to source distance, the spot size, the source temporal stability;
 \item the spectral response as a function of energy;
 \item the map of the gain over the detector sensitive area;
 \item the detector dead time as a function of energy.
\end{itemize}

While results of calibration measurements listed above will be reported elsewhere, here we stress that each calibration required a specific source, either polarized, unpolarized or of both kinds. Each source was assembled, tested with a commercial spectrometer and a commercial imager, and eventually aligned with the DU before the calibration measurement. This could take a few hours for the most demanding configurations and, therefore, the sequence of calibration measurements was chosen to minimize the number of switches in the set-up.

Measurements could last from minutes to several hours. As the detector and the sources could be autonomously and safely switched off in case of power failure, we carried out longer measurements during the night, when the on-going measurement was monitored remotely. This allowed us to acquire data essentially 24 hours per day and 7 days per week during the whole $>$6~months-long calibration campaign, with breaks just during the source switches.

Calibration, especially the study of the response to polarized and unpolarized radiation, covered all the sensitive area of the focal plane detector, which is 15$\times$15~mm$^2$, with the so-named \emph{Flat Field} (FF) measurements. However, these measurements were also repeated with higher sensitivity in the central region with 3.3~mm radius, which is named \emph{Deep Flat Field} (DFF). Such a strategy descends from the one chosen for in-orbit observations, which will be based on the dithering of the target over a circular region with the radius of $\approx$1.5~mm. The brightest sources, for which IXPE will achieve higher sensitivity, will be point-like and, once accounting for the mutual misalignment among telescopes and for the pointing error of the satellite, they will dither in the central, 3~mm radius, area for all the three DUs. Therefore, the latter region is calibrated with higher accuracy.

All the efforts were spent to configure the sources so as to obtain a spectrum largely dominated by photons at a single energy. This facilitates the subsequent data analysis, as it is easier with nearly monochromatic photons to deconvolve the energy-dependent characteristics of the detector. When not limited by the source flux, a counting rate of $\lesssim$200~cts/s is used for all the measurements, which corresponds essentially to the counting rate expected from an astronomical source as bright as the Crab Nebula.

During the test of the Instrument, that is, of the three DUs integrated with the DSU, the aim was to stress the Instrument and its interfaces at the maximum rate of $\sim$300~cts/s, verifying that scientific performance was unaffected. In this phase, we used three of the sources designed for calibration to illuminate contemporaneously the three DUs interfaced with the DSU. Sources were chosen preferentially to provide an average energy close to the peak of sensitivity of the DU ($\sim$3~keV).

\begin{figure*}[t]
	\begin{center}  
		\includegraphics[width=14cm]{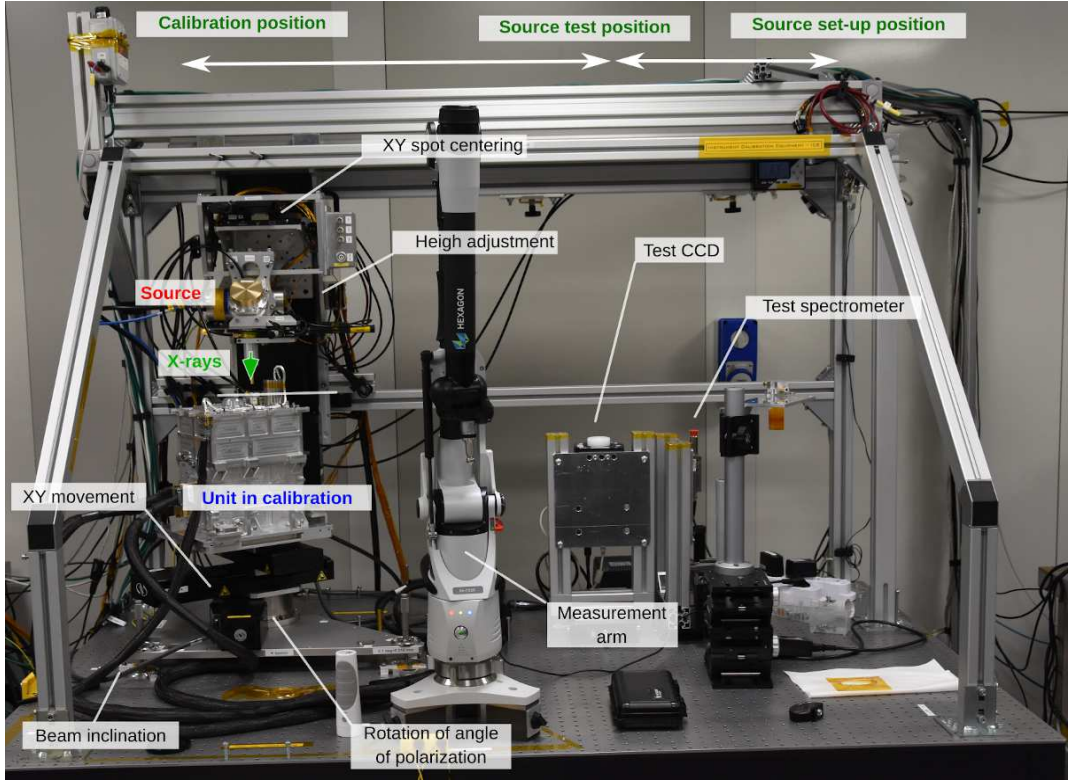}
	\end{center}
	\caption{Overview of the ICE. The measurement arm used for alignment, see Section~\ref{sec:Alignment}, is at center. Two commercial detectors (on the right) are available to test the source before the calibration of the DU.}
	\label{fig:ICE_1} 
\end{figure*}

\section{The ICE and the ACE}

Two independent apparatuses were built for IXPE Instrument calibration and testing, one is dedicated to DU calibration and named \emph{Instrument Calibration Equipment}, or ICE, and one is named \emph{AIV-T Calibration Equipment}, or ACE (see Figure~\ref{fig:ICE_ACE}). The ACE was initially designed for allowing the contemporaneous illumination of up to three DUs during the assembly, integration and test phase of the integrated Instrument, but it was soon upgraded to also carry out specific calibration measurements. In fact, the same sources can be mounted on both the ICE and the ACE. However, the ICE offers a full-fledged set of manual and motorized stages to align and move the source beam in a controlled way even during the measurement, whereas the ACE was equipped only with a subset of manual stages. Nevertheless, these provided the capability to perform also on ACE, and contemporaneously with the ICE, the most time-consuming calibrations, that are, the calibration of the response to unpolarized radiation. 

\begin{figure*}
\begin{center}
\begin{tabular}{c}
\subfloat[\label{fig:tower_assembly}]{\includegraphics[width=8cm]{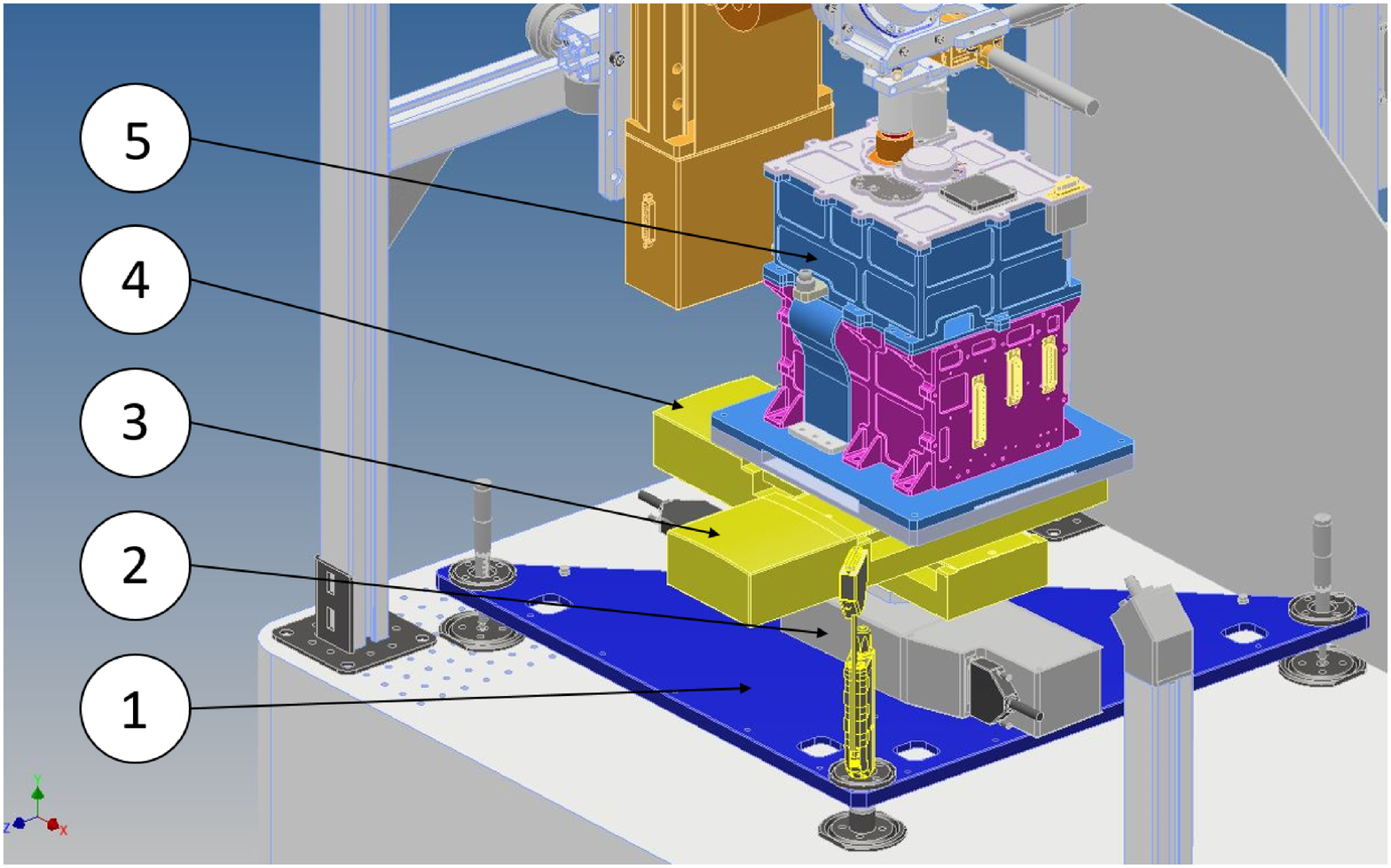}}
\hspace{0.5cm}
\subfloat[\label{fig:source_assembly}]{\includegraphics[width=8cm]{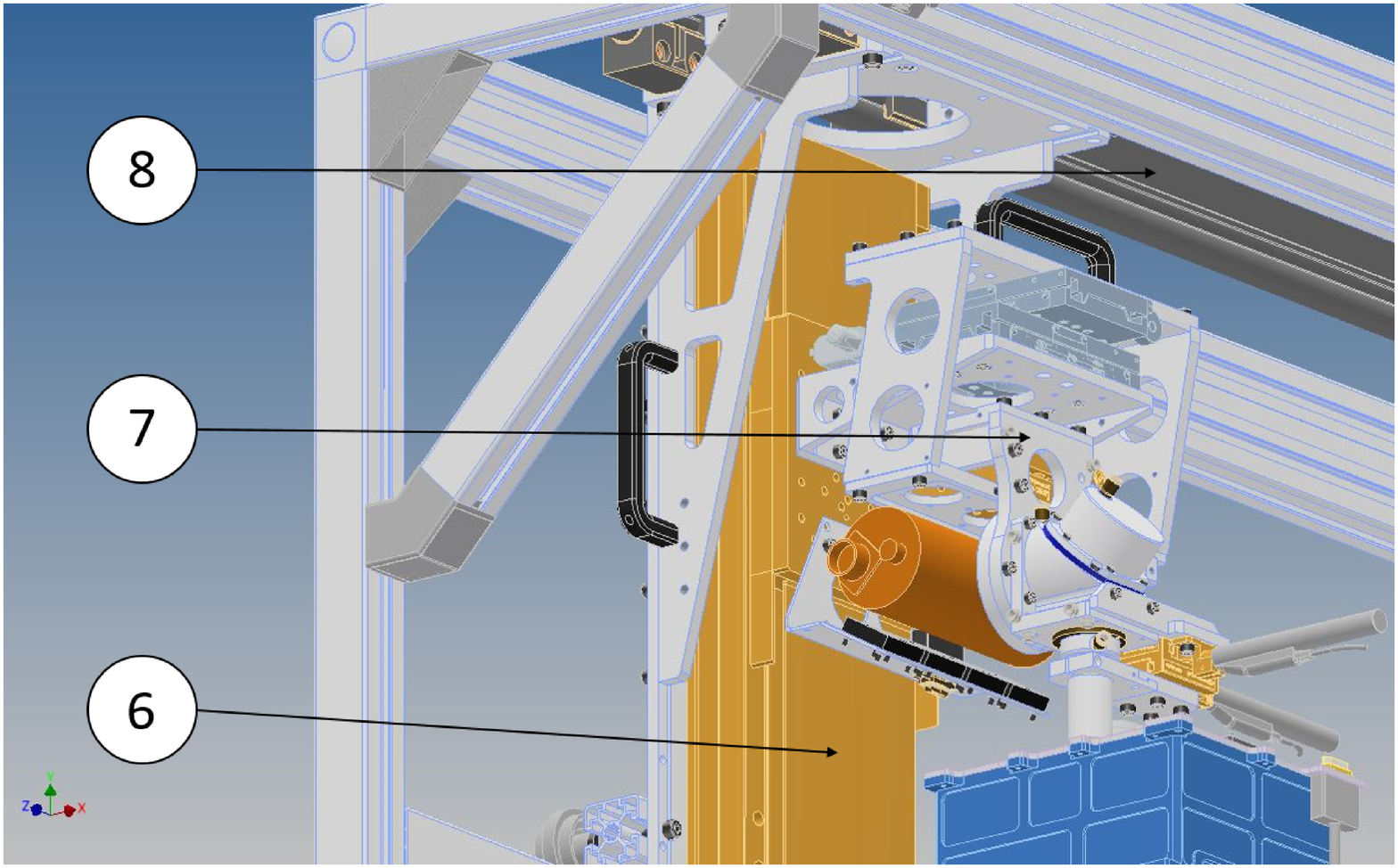}}
\end{tabular}
\end{center}
\caption{Detailed view of the tower supporting the DU (a) and of the source assembly (b). The items in the figures are the manual and motorized stages which allow one to align the X-ray beam and the DU (item 5). The beam is centered on the detector with two linear stages, item 3 and 4, and its angle of polarization with respect to the detector is changed with a third rotation stage (item 3). The detector tip/tilt is adjusted with a platform (item 1). The source height and tip/tilt are changed with item 6 and 7, respectively, and its position is slide with a manual stage (item 8).}
\end{figure*}

A picture of the ICE is shown in Figure~\ref{fig:ICE_1}. A set of manual and motorized stages permit adjustment of the beam direction with respect to the unit in calibration. There are two groups of stages which are controlled independently: the first is named \emph{ALIGN} and it is dedicated to aligning the source to the DU before the actual calibration. The second set of stages, named \emph{MEAS}, is dedicated to moving the DU with respect to the source during calibration to, e.g., sample the field of view of the instrument. \emph{MEAS} stages are assembled on the tower on which the DU is mounted and they allow to (see Figure~\ref{fig:tower_assembly}):
\begin{itemize}
 \item move the DU (item 5 in the figure) on the plane orthogonal to the incident beam with an accuracy of $\pm$2~$\mu$m (over a range of 100~mm) to move the beam across the detector sensitive surface. These two stages (items 4 and 3, model Newport ILS-100CC) are named \emph{xdu} and \emph{ydu}.
 \item rotate the DU on the azimuthal plane which is orthogonal to the incident beam with an accuracy of $\pm$7~arcsec, to test the response at different polarization angle and to average residual polarization of unpolarized sources. This stage (item 2, model Newport RV-120CC) is named $\epsilon$.
 \item tip/tilt (item 1) the DU to align it to the incident beam. Two out of the three actuators of the tip/tilt plate (called $\eta_1$ and $\eta_2$) are manual micrometers, but one ($\eta_0$) is motorized to have the possibility to carry out automatically measurements with the beam off-axis at a series of known angles, between $<$1 degree and about 5 degrees, e.g., to simulate the focusing of X-ray mirror shells. 
\end{itemize}

The X-ray source is mounted on a mechanical support to adjust its position and inclination with respect to the DU (see Figure~\ref{fig:source_assembly}) through the \emph{ALIGN} group of stages. A manual translation stage ($\nu$, item 8) slides the source assembly to three separate positions to perform: (1) the calibration with the DU; (2) the test of the source with commercial detectors; (3) the source set-up and mounting (see Figure~\ref{fig:ICE_1}). A vertical motorized stage (\emph{zso}, model Newport IMS300V, item 6) with range of 300~mm and accuracy of $\pm$5~$\mu$m lifts the source to different heights and minimize the distance of the source to the DU. A couple of motorized linear stages (models Newport M-436A, item 7) mounted in XY configuration on the plate of the vertical stage, named \emph{xso} and \emph{yso}, moves the X-ray source to center the beam to the axis of rotation of $\epsilon$, to avoid moving the spot while rotating this stage.

\begin{figure}[t]
\begin{center}
\begin{tabular}{c}
\includegraphics[width=8cm]{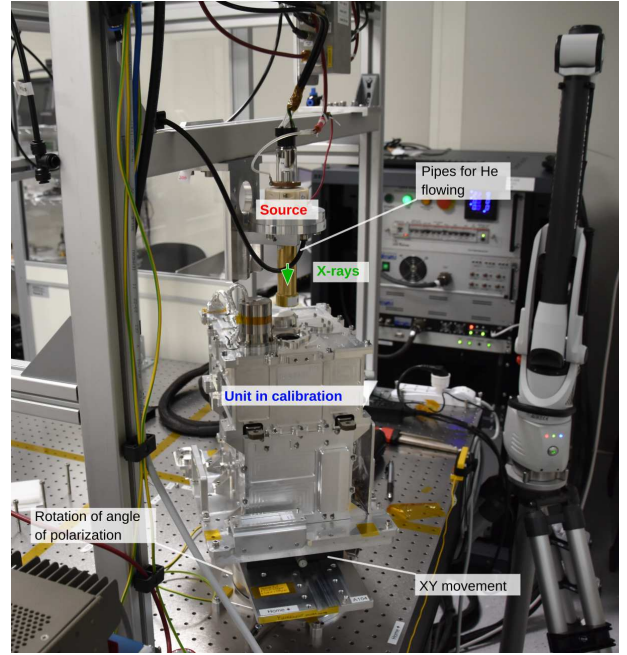}
\end{tabular}
\end{center}
\caption{Picture of the ACE with DU-FM2.}
\label{fig:ACE} 
\end{figure}

A picture of the ACE is shown in Figure~\ref{fig:ACE}. The source is mounted on a frame which allows tip/tilt adjustment of the beam with shims. X-ray spot centering and rotation with respect to the DU is possible with the manual stages mounted below the unit.

Calibration sources are tested and characterized before each measurement with the DU. Commercial X-ray detectors are available for this purpose in the ICE (see Figure~\ref{fig:ICE_1}):
\begin{itemize}
 \item a CCD imager with 1024$\times$1024 13-$\mu$m pixels, model Andor iKon-M SY. This is used to image the beam spot.
 \item a SDD spectrometer and photometer model Amptek FAST SDD 7$\times$7mm$^2$. This is used to characterize the spectrum and measure the flux of the source. It is also the reference detector for the efficiency measurements of the DU efficiency.
\end{itemize}

\section{Unpolarized calibration sources}\label{sec:Unpolarized}

\begin{figure}
\begin{center}
\begin{tabular}{c}
\includegraphics[width=8cm]{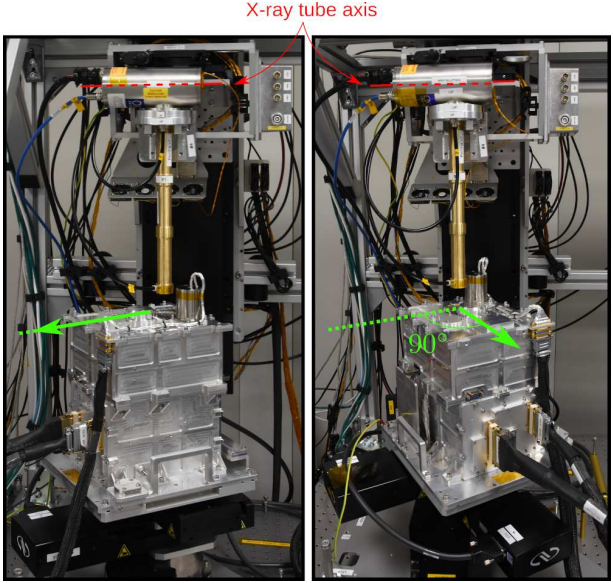}
\end{tabular}
\end{center}
\caption{Set-up of the two measurements carried out for decoupling the intrinsic instrumental response to unpolarized radiation from the signal due to the genuine source polarization.}
\label{fig:decoupling} 
\end{figure}

Unpolarized sources in the ICE are of different kinds to cover the entire energy range of IXPE detector. Only in a few cases it is possible to use ``genuine'' unpolarized sources, in the sense that any source polarization is much lower than the statistical significance achieved in the calibration. Therefore, we adopted a simple procedure to decouple and measure contemporaneously the source intrinsic polarization and the response of the instrument to completely unpolarized radiation. Such a procedure, described in detail in \cite{Rankin2021}, is based on the repetition of the measurement with the same source at two different azimuthal angles, typically rotated of 90$^\circ$ one with respect to the other (see Figure~\ref{fig:decoupling}). In the two measurements, the signal due to the real polarization of the source and the intrinsic instrumental response combine differently, as only the former component rotates according to the change of the azimuthal angle, which is known. Therefore, the two measured values of polarization degree and angle, or, equivalently the Stokes parameters, provide the four quantities from which it is possible to derived the four unknown values which characterize the amplitude and phase of the modulation due to the real source polarization and to the intrinsic instrumental effects.

Both radioactive sources and X-ray tubes are used. Radioactive nuclei emit in general unpolarized radiation, but in the IXPE energy range only $^{55}$Fe can be used. This nuclide has half-life of 2.737~years and it decays into $^{55}$Mn for K-capture with emission of Mn fluorescence lines K$\alpha$ (5.90 keV, 90\%) and K$\beta$ (6.49 keV, 10\%). Both lines are expected to be unpolarized from first principle and no hint of polarization has indeed been found. The source polarization, obtained as a by-product of the calibration of the DU response to unpolarized radiation, was lower than the Minimum Detectable Polarization (MDP) of the measurement (see Table~\ref{tab:UnpolSource}). The MDP is the maximum polarization which can be attributed to the statistical fluctuations in the measurement at a 99\% confidence level \cite{Weisskopf2010}, and therefore only when the measured value is higher than the MDP it is statistically significative. The activity of the source used for DU calibration is about 4 mCi.

\begin{figure}
\begin{center}
\begin{tabular}{c}
\includegraphics[width=7cm]{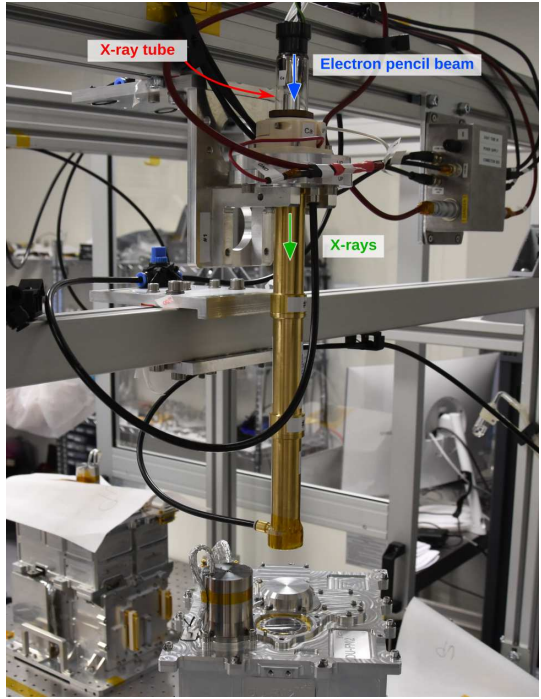}
\end{tabular}
\end{center}
\caption{Geometry of head-on X-ray tube with Calcium anode.}
\label{fig:HeadOnXrayTube} 
\end{figure}

\begin{figure}
\begin{center}
\begin{tabular}{c}
\includegraphics[width=8cm]{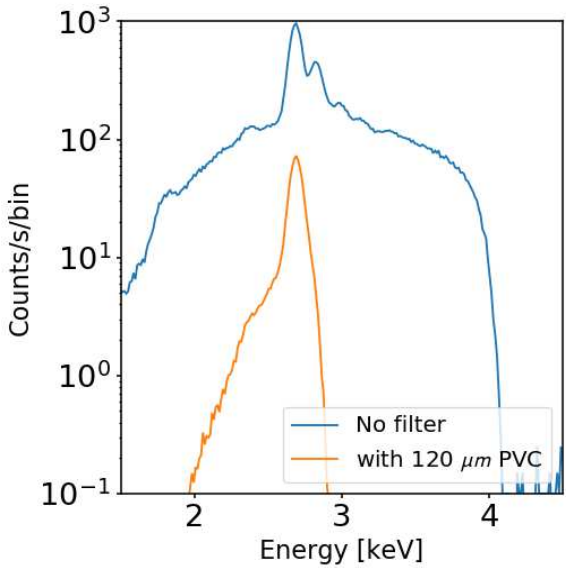}
\end{tabular}
\end{center}
\caption{Example of the effect on the spectrum of direct X-ray tube with Rh anode with and without X-ray filter 120~$\mu$m thick to select the L$\alpha$ fluorescence line.}
\label{fig:filter} 
\end{figure}

Measurements at other energies are carried out with X-ray tubes in two different configurations, one direct and one for the extraction of fluorescence emission from a target. In the direct configuration, the DU is illuminated with the direct emission of the X-ray tube (or $^{55}$Fe), which comprises unpolarized fluorescence lines and continuum bremsstrahlung emission. The latter may be partially polarized depending on the details of the X-ray tube emission geometry. In particular, two different geometries are used:
\begin{itemize}
 \item Head on X-ray tube with Calcium anode by Hamamatsu, model N1335. The spectrum includes the Ca~K lines (K$\alpha$ at 3.69~keV and K$\beta$ at 4.01~keV) plus bremsstrahlung. The geometry of this tube is such that X-ray photons are generated in the same direction as the electron pencil beam which hits the Ca target (see Figure~\ref{fig:HeadOnXrayTube}). In this case, the polarization of the brems\-strah\-lung emission is very low for symmetry reasons and a null polarization is expected including also the fluorescence emission. This is confirmed by the value measured during DU calibration (see Table~\ref{tab:UnpolSource}).
 \item Right-angle X-ray tubes with Rh or Ag anodes by Oxford series~5000. The spectrum of each X-ray tube comprises the fluorescence lines of its material, plus bremsstrahlung which in this case is polarized to few tens of \% and increasing with energy \cite{Ratheesh2021}. X-ray filters are used to suppress the continuum and polarized bremsstrahlung emission with respect to the unpolarized fluorescence radiation, thus reducing the intrisic source polarization. For example, the X-ray tube with rhodium anode emits L$\alpha$ and L$\beta$ fluorescence at 2.70 and 2.84~keV. A filter made of polyvinyl chloride can absorb most of bremsstrahlung above 2.82~keV (and the L$\beta$ fluorescence), which correspond to the K-shell absorption energy of chlorine (see Figure~\ref{fig:filter}). The source polarization, including the X-ray filters, is measured during calibration and it is 6.7\% at 2.7~keV and 13.2\% at 2.98 keV (see Table~\ref{tab:UnpolSource}). Interestingly enough, the polarization direction is parallel to the axis of symmetry of the cylindrical package of the X-ray tube (see Figure~\ref{fig:decoupling}), that is, it is aligned with the direction of the tube electron beam.
\end{itemize} 

\begin{figure}
\begin{center}
\begin{tabular}{c}
\subfloat[\label{fig:unpol_direct_draw}]{\includegraphics[width=9cm]{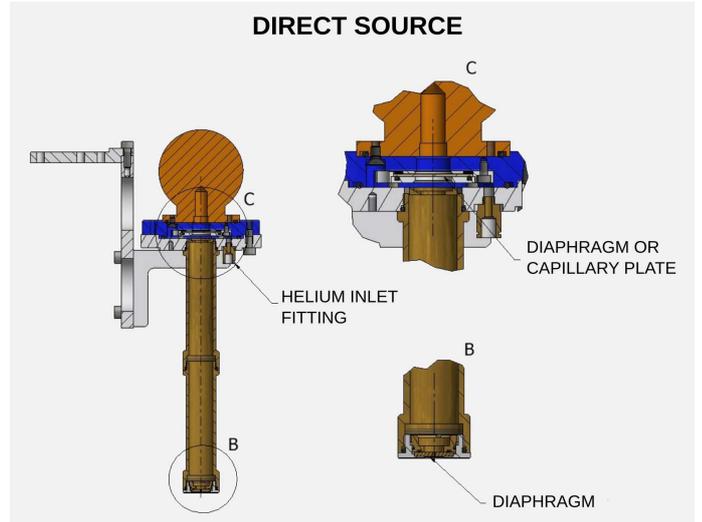}}
\\
\subfloat[\label{fig:unpol_flourescence_draw}]{\includegraphics[width=9cm]{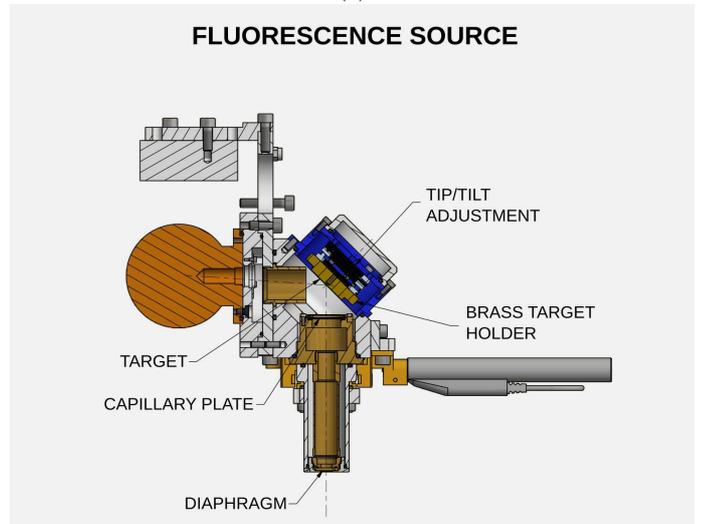}}
\end{tabular}
\end{center} 
\caption{Drawing of the direct unpolarized source (a) and of the fluorescence unpolarized source (b).}
\end{figure}

A drawing of the direct unpolarized source with an Oxford right-angle X-ray tube is shown in Figure~\ref{fig:unpol_direct_draw}. Two diaphragms, one close to the X-ray spot and one at the other end, can be used to collimate the beam and constrain the beam direction. Collimator length is adjustable to make a trade-off between the source counting rate and the aperture of the beam and it is made of brass. This material is chosen as it is easy to procure and machine; moreover its intermediate (average) atomic number favors photoelectric absorption over scattering in the IXPE energy band and its fluorescence lines are outside such an energy interval. This reduces the reprocessing of the primary emission of the X-ray tube by the collimator and ultimately decreases the generation of background photons in the working energy band, at the cost of a relatively-low increase in the collimator mass with respect to one made of, e.g., aluminum. The assembly is air-tight to allow helium flowing and reduce air absorption (see Section~\ref{sec:Absorption}). Total distance from the X-ray spot to the lower collimator is 344~mm. 

\begin{figure}
\begin{center}
\begin{tabular}{c}
\includegraphics[totalheight=4.3cm]{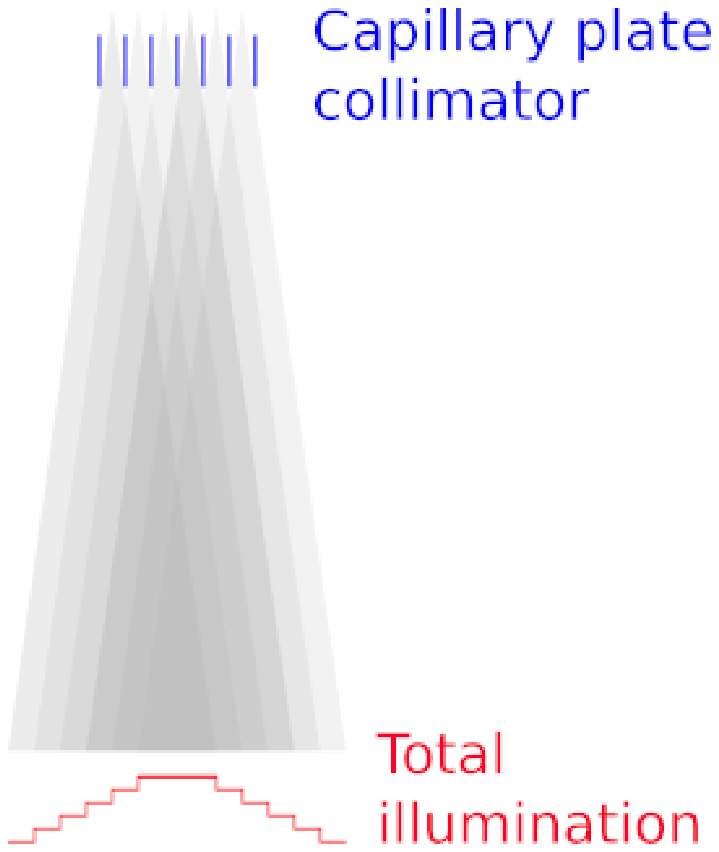}
\includegraphics[totalheight=4.3cm]{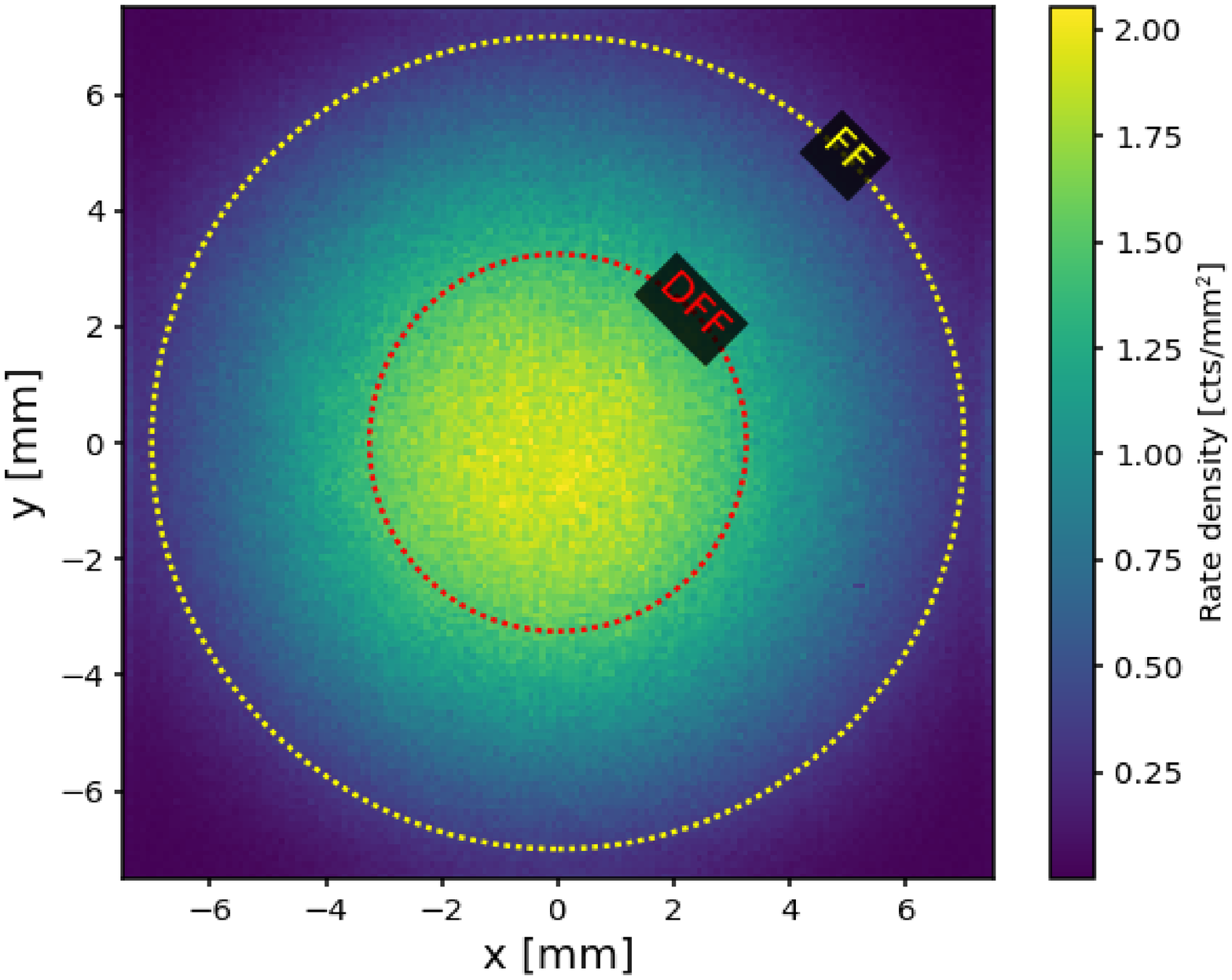}
\end{tabular}
\end{center}
\caption{(Left) Illustration of the geometrical effect which causes a non-uniform illumination for fluorescence sources. (Right) Illumination measured for fluorescence source at 2.29~keV.}
\label{fig:fluorescence} 
\end{figure}

Fluorescence unpolarized sources are used to produce 2.04 and 2.29~keV X-rays. A drawing of the source is shown in Figure~\ref{fig:unpol_flourescence_draw}; this is the polarized source described in Section~\ref{sec:Polarized} but with the crystal replaced with the fluorescence target. Different materials can be used as target; we used Zirconium, whose L$\alpha$ and L$\beta$ fluorescence are at 2.04~keV and 2.12~keV, respectively, and, Molybdenum, whose L$\alpha$ and L$\beta$ fluorescence are at 2.29~keV and 2.39~keV, respectively. Since fluorescence is emitted on an extended region and it is isotropic, a capillary plate collimator with collimation of $\pm$1.4$^\circ$ is used to obtain a nearly parallel and extended beam incident on the detector. However, source illumination is not uniform because of the combined effect of finite size of the source, not-ideal capillary plate collimation and distance between the capillary-plate collimator and the detector sensitive area (see Figure~\ref{fig:fluorescence}). Calibration time is therefore adjusted to collect the required statistics also at the edge of the Flat Field (FF) region.

\begin{figure}
\begin{center}
\begin{tabular}{c}
\includegraphics[width=8cm]{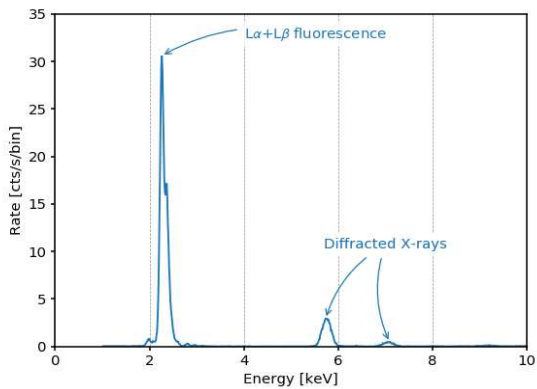}
\end{tabular}
\end{center}
\caption{Spectrum of Zr fluorescence source measured with ICE test spectrometer.}
\label{fig:fluorescenceSpectrum} 
\end{figure}

\begin{table*}
 \caption{Summary table for unpolarized sources available in the ICE. Reported values were measured with the fourth DU flight model; values from other DUs are similar. }
 \label{tab:UnpolSource} 
 \begin{center}
  \begin{tabular}{c|c|c|c|c|c}
   Energy & Configuration & Source & Rate & \multicolumn{2}{c}{Polarization }\\
   & & settings & [cts/s] & Degree [\%] & Angle [deg] \\
   \hline
   \multirow{2}{*}{2.04} & Fluorescence of Zr target & \multirow{2}{*}{19~kV, 0.95 ~mA} & \multirow{2}{*}{$\sim$177} & \multirow{2}{*}{1.12$\pm$0.14} & \multirow{2}{*}{1.38$\pm$3.62\dag} \\
        & illuminated by Rh X-ray tube & & & \\
   \hline
   \multirow{2}{*}{2.29} & Fluorescence of Mo target & \multirow{2}{*}{17~kV, 0.77~mA} & \multirow{2}{*}{$\sim$199} & \multirow{2}{*}{0.85$\pm$0.11} & \multirow{2}{*}{1.32$\pm$3.63\dag} \\
        & illuminated by Ag X-ray tube & & & \\
   \hline
   \multirow{2}{*}{2.70} & Direct X-ray tube with & \multirow{2}{*}{4.0~kV, 0.52mA}  & \multirow{2}{*}{$\sim$205} & \multirow{2}{*}{6.71$\pm$0.10} & \multirow{2}{*}{-0.93$\pm$0.41\dag}\\
        & Rh anode + 300~$\mu$m PVC filter & & & \\
   \hline
   \multirow{2}{*}{2.98} & Direct X-ray tube with & \multirow{2}{*}{4.0~kV, 0.06~mA}  & \multirow{2}{*}{$\sim$215} & \multirow{2}{*}{13.20$\pm$0.09} & \multirow{2}{*}{-0.19$\pm$0.19\dag}\\
        & Ag anode + 6~$\mu$m Ag filter & & & \\
   \hline
   \multirow{2}{*}{3.69} & \multirow{2}{*}{Direct X-ray tube with Ca anode} & \multirow{2}{*}{5.4~kV, 0.003~mA} & \multirow{2}{*}{$\sim$213} & Undetected, & \multirow{2}{*}{Undetected}\\
   & & & & MDP(99\%)=0.22\% & \\
   \hline
   \multirow{2}{*}{5.89} & \multirow{2}{*}{$^{55}$Fe nuclide} & \multirow{2}{*}{4mCi} & \multirow{2}{*}{$\sim$180} &  Undetected, & \multirow{2}{*}{Undetected}\\
   & & & & MDP(99\%)=0.19\% \\ 
  \end{tabular}
 \end{center}
 \dag Polarization angle is measured with respect to the axis of the package of the X-ray tube, see Figure~\ref{fig:decoupling}.
\end{table*}

It is worth noting that the spectrum of fluorescence sources has different components (see Figure~\ref{fig:fluorescenceSpectrum}). In addition to the prominent unpolarized fluorescence emission, lines at higher energy are present. We associated the latter to the Bragg diffraction on the fluorescence target. This interpretation is supported by two pieces of evidence: (i) the energy of the line changes when the inclination of the target is changed and (ii) that the higher energy photons are highly polarized, about 44\%. We limited the presence of this component by mounting slightly off-axis the X-ray tube spot, so that diffracted photons were partially blocked by the collimator. The residual contribution was not an issue for DU calibration. On the one hand, spectral capabilities of the DU are sufficient to remove the large majority of this component; on the other hand, any residual contribution would be identified as a contribution of the source and decoupled from the intrinsic response of the instrument to unpolarized radiation. As a matter of fact, the intrinsic polarization of fluorescence sources is measured to be very small once the fluorescence line is selected in energy (see Table~\ref{tab:UnpolSource}), with an angle of polarization aligned with the axis of the X-ray tube (see Figure~\ref{fig:decoupling}). This is expected as in Bragg diffraction polarization angle is orthogonal to the diffraction plane, which happens to be roughly aligned with the tube axis.

\begin{figure}
\begin{center}
\begin{tabular}{c}
\includegraphics[width=8cm]{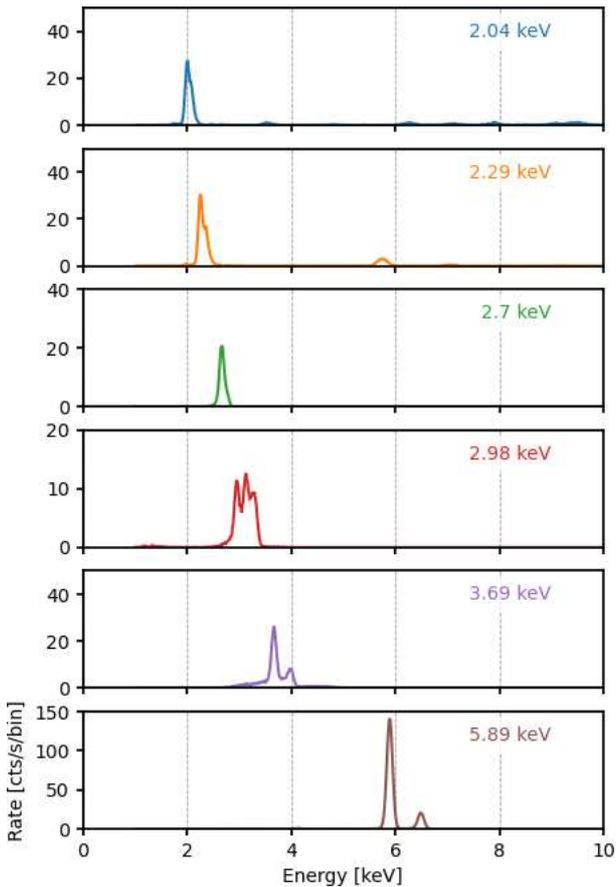}
\end{tabular}
\end{center}
\caption{Spectra of unpolarized sources measured with the ICE test spectrometer. X-ray tube high-voltage setting and distance to detector are equivalent to those used for DU calibration.}
\label{fig:unpolSpectra} 
\end{figure}

A summary of the X-ray tube configurations used for DFF unpolarized radiation calibration is in Table~\ref{tab:UnpolSource}. Reported values are those obtained with the fourth DU flight model; values measured with other DUs are similar. Spectra of each source, measured with the ICE test spectrometer in conditions equivalent to DU calibration (same X-ray tube high-voltage setting, equivalent source-detector distance and absorption) are shown in Figure~\ref{fig:unpolSpectra}.

\section{Polarized sources} \label{sec:Polarized} 

\begin{figure*}
\begin{center}
\begin{tabular}{c}
\includegraphics[width=14cm]{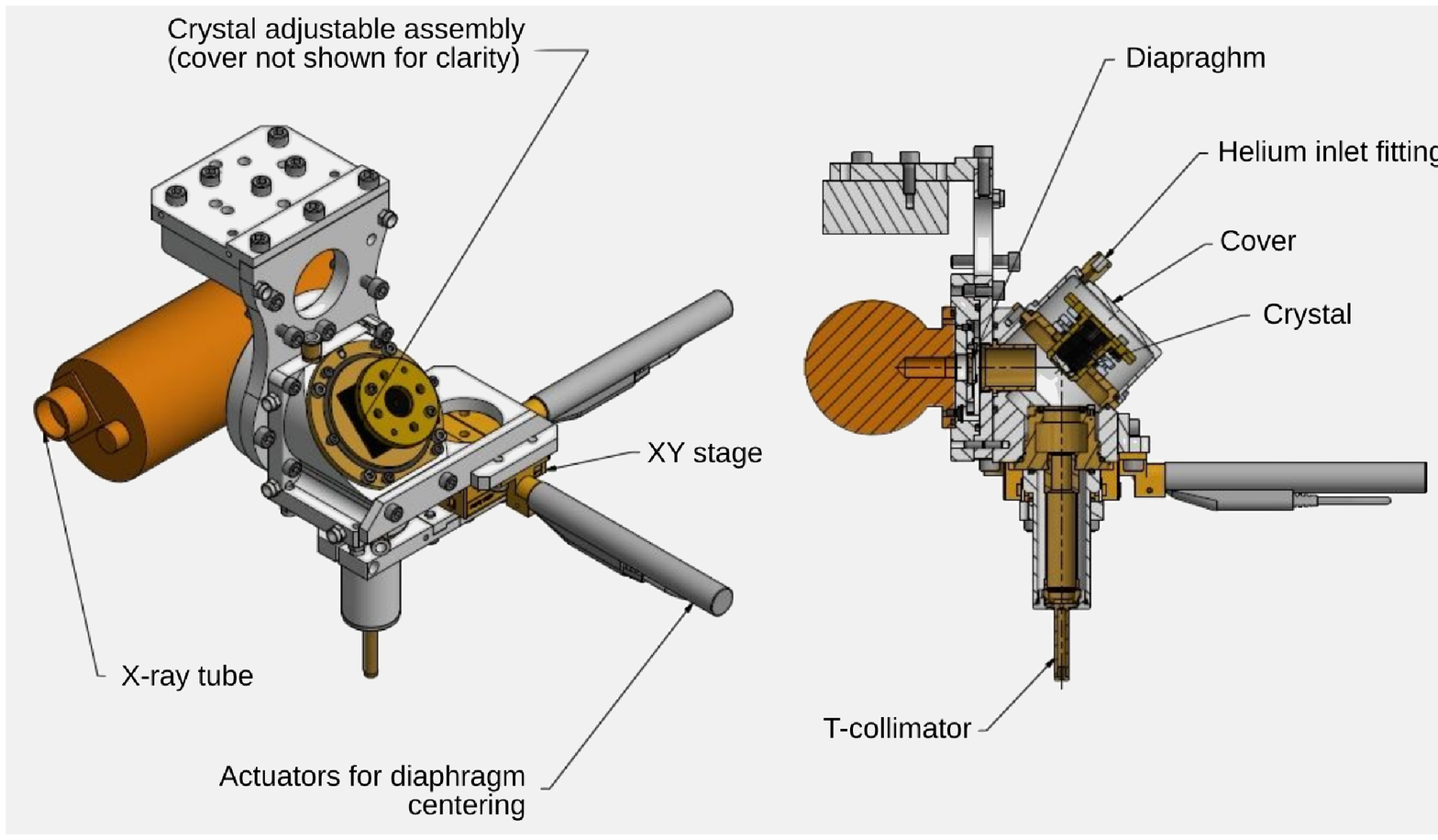}
\end{tabular}
\end{center}
\caption{Drawing (top) and cut-view (bottom) of the ICE polarized source.}
\label{fig:pol_draw} 
\end{figure*}

\begin{figure}
\begin{center}
\begin{tabular}{c}
\includegraphics[width=8cm]{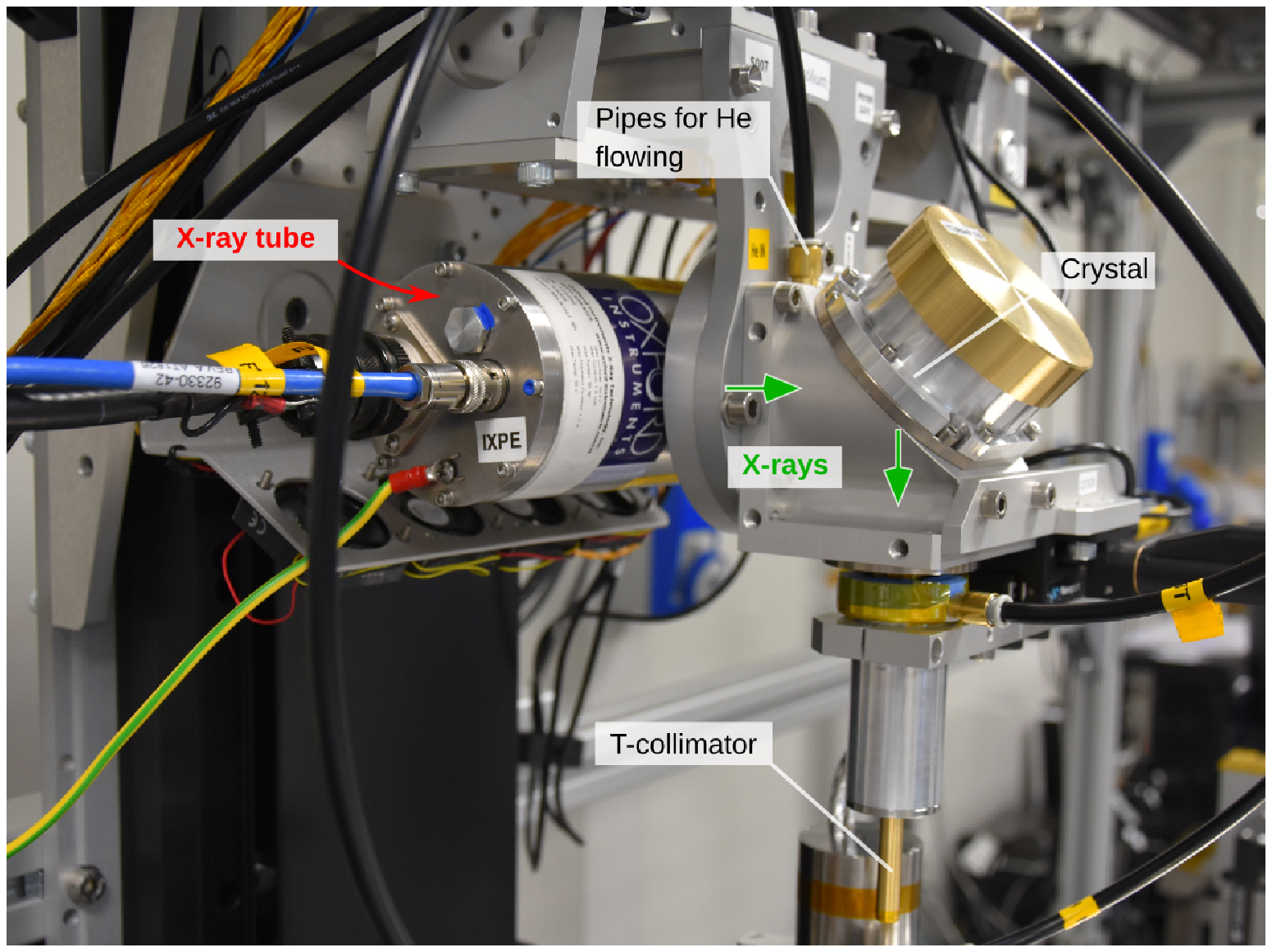}
\end{tabular}
\end{center}
\caption{ICE polarized source during calibration of DU-FM2 with 2.70 keV polarized X-rays.}
\label{fig:PolSource} 
\end{figure}

ICE polarized sources are based on Bragg diffraction at nearly 45~degrees, with a design based on the heritage of the calibration facility used at INAF-IAPS for GPD characterization for more than 10~years \cite{Muleri2007}. X-rays are generated with a commercial X-ray tube (Oxford Series~5000 or Hamamatsu Head-on~N7599 series) and then diffracted at nearly 45~degrees from a crystal (see Figure~\ref{fig:PolSource}). A different tube and crystal pair is used at each energy, and for each of them the diffraction angle is chosen so that the Bragg energy for the crystal matches the energy of the most prominent fluorescence line produced by the X-ray tube. In this condition, the input radiation can be assumed to be essentially monochromatic at the energy of the fluorescence line, and this precisely and uniquely determines the diffraction angle and then the polarization of the diffracted radiation. The latter is calculated from the reflectivity values found in the literature \cite{Henke1993}. When fluorescence lines are not available at the energy of interest, continuum radiation from the X-ray tube is diffracted and the incident and diffraction angles are tightly constrained around 45~degrees to have nearly 100\%\--polarized X-rays. 
   
A drawing of ICE polarized source is in Figure~\ref{fig:pol_draw}. Collimators can be mounted to constrain the incident and/or diffracted direction of X-rays. Actual configuration depends on many factors, e.g., crystal reflectivity, ratio of fluorescence line to continuum in the X-ray tube spectrum, etc, and is aimed to have a clean diffracted spectrum and a sufficiently high flux. The crystal is integrated with a Newport MFM-100 tip-tilt stage which was used to align it to input and/or output collimator to better than 0.03~deg (or 1.8~arcmin). Diaphragms with diameter starting from 25~$\mu$m can be mounted in front of the X-ray tube spot to limit X-rays scattering inside the source or on the diffracted beam to reduce its size. In the latter configuration, the diaphragm is centered with the beam thanks to two Newport 460A-XY linear stages actuated by Newport TRB12CC motorized actuators. The source is made of aluminum to reduce its mass, but its inner parts along the X-ray path are made of brass to reduce the reprocessing of primary emission from the X-ray tube and reduce the source background. The assembly is air-tight to allow helium flowing (see Section~\ref{sec:Absorption}).

\begin{table*}
 \caption{Configurations of polarized sources. Crystals in italic were tested but eventually discarded for IXPE Instrument calibration. Polarization degree is calculated from \cite{Henke1993}. Polarization angle is orthogonal to the diffraction plane and it is measured as the intersection line between the crystal plane and the detector plane (see Section~\ref{sec:Alignment}).} 
 \label{tab:crystal} 
 \begin{center}    
 \begin{tabular}{c|c|c|c|c|c|c|c}
  Crystal & Incident  & Energy  & Diffraction & Polarization & X-ray tube & Rate \\
          & radiation &         & angle       &              & settings   & \\
          &           & [keV]   & [deg]       & [\%]         &            & [cts/s] \\
  \hline
  PET (002) & Continuum              & 2.01 & 45.00 & $\approx$100\% & 20~kV, 0.8~mA  & $\sim$30  
\\
  \hline
  \emph{ADP (200)} & Mo L$\alpha$    & 2.29 & 46.13 & 99.5\%          &                &           \\
  InSb (111) & Mo L$\alpha$          & 2.29 & 46.28 & 99.3\%          & 17~kV, 0.06~mA & $\sim$150 \\
  \hline
  \emph{Graphite (0002)} & \emph{Continuum} & \emph{2.61} & \emph{45.00} & \emph{$\approx$100\%} &  
& \\
  Ge (111) & Rh L$\alpha$            & 2.70 & 44.73 & 99.5\%          & 17~kV, 0.03~mA & $\sim$150 \\
  \hline
  Si (111) & Ag L$\alpha$            & 2.98 & 41.49 & 95.1\%          & 17~kV, 0.2~mA  & $\sim$140 \\
  \hline
  Al (111) & Ca K$\alpha$            & 3.69 & 45.90 & 99.4\%          & 10~kV, 0.01~mA & $\sim$35  \\
  \hline
  \emph{CaF$_2$ (220)} & \emph{Ti K$\alpha$} & \emph{4.51} & \emph{45.40} & \emph{99.9\%} & &  \\
  Si (220) & Ti K$\alpha$            & 4.51 & 45.73 & 99.5\%          & 18~kV, 0.25~mA & $\sim$160 \\
  \hline
  \emph{LiF (220)} & \emph{Fe K$\alpha$} & \emph{6.40} & \emph{42.86} & \emph{90.0\%} & & \\
  Si (400) & Fe K$\alpha$            & 6.40 & 45.51 & $\approx$100\% & 25~kV, 0.85~mA & $\sim$150\\
 \end{tabular}
 \end{center}    
\end{table*}

\begin{figure}[t]
\begin{center} 
\begin{tabular}{c}
\includegraphics[width=9cm]{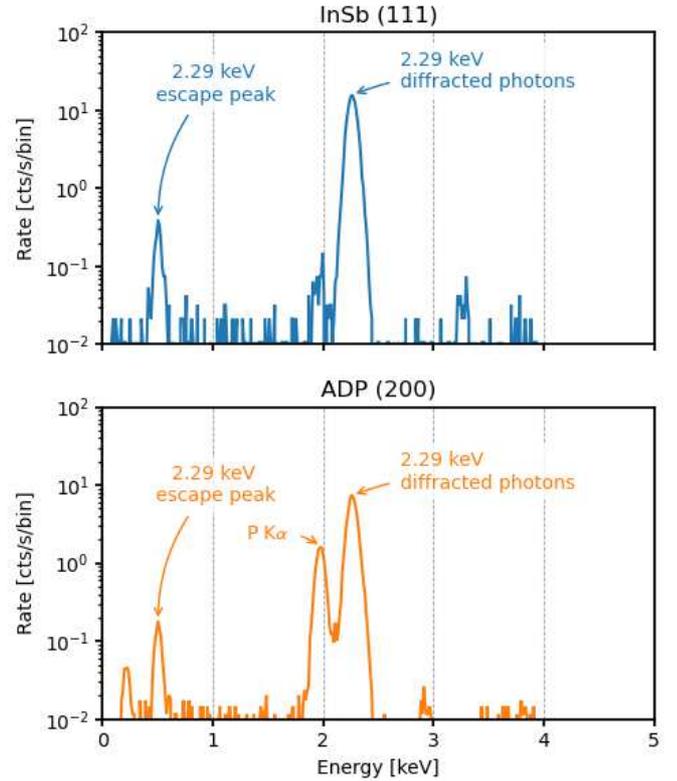}
\end{tabular}
\end{center}
\caption{Comparison of spectra measured with the ICE test spectrometer of the polarized source at 2.29~keV with InSb~(111) or ADP~(200) crystals for the same X-ray tube settings. For the latter, the counting rate is lower and the presence of the fluorescence lines of phosphorus makes it difficult to estimate the polarization of the combined lines, which would be unresolved with the spectral resolution of the DU.}
\label{fig:ADPspectrum} 
\end{figure}

Different crystals, with different lattice spacing, were procured to diffract X-rays over the whole DU energy range (see Table~\ref{tab:crystal}). In some cases, more than one crystal was available at the same energy; then, all the crystals were tested and the one actually used for DU calibration was selected according to the following criteria:
\begin{itemize}
 \item InSb~(111) crystal was preferred to ADP~(200) to diffract 2.29~keV because the former has a higher reflectivity, resulting in a higher counting rate. Moreover, ADP contains phosphorus, which has prominent K-fluorescence (and unpolarized) lines at 2.01~keV (see Figure~\ref{fig:ADPspectrum}). Such lines would not have been resolved with the spectral capabilities of the DU and this would have made the determination of the source polarization difficult.
 \item Graphite was not used because a setup with a Ge~(111) crystal diffracting L$\alpha$ fluorescence of rhodium provided a much cleaner spectrum and a much higher counting rate basically at the same energy.
 \item Si~(220) was preferred to CaF$_2$~(220) because the Calcium fluorescence at 3.69 and 4.01~keV from the crystal would not have been resolved from the diffracted 4.51~keV photons with the detector spectral resolution.
\end{itemize}

\begin{figure}
\begin{center}
\begin{tabular}{c}
\includegraphics[width=7cm]{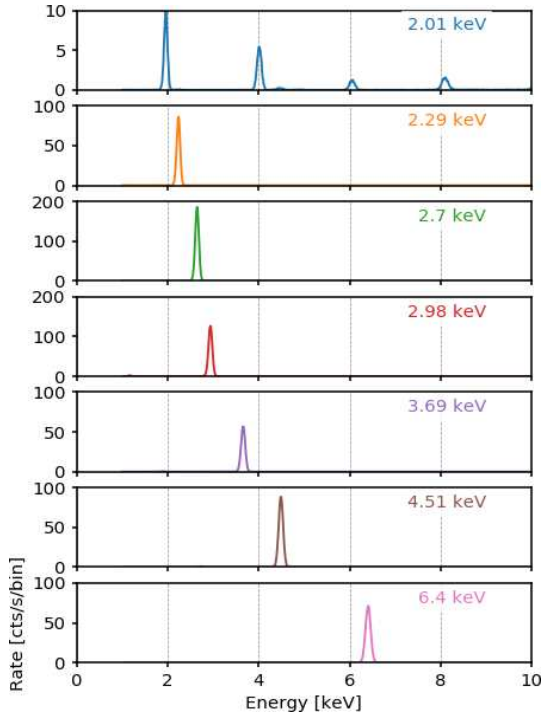}
\end{tabular}
\end{center}
\caption{Spectra from polarized sources measured with the ICE test spectrometer.}
\label{fig:polSpectra} 
\end{figure}

Spectra of sources used for the calibration of the DU response to polarized radiation are presented in Figure~\ref{fig:polSpectra}. When the fluorescence emission of the X-ray tube is tuned with the crystal, that is, in all cases except 2.01~keV, the spectrum is very clean, with only a prominent diffracted line. In the case of the source at 2.01~keV, continuum radiation is diffracted and higher orders of diffraction have a comparable intensity with respect to the first order which is used for calibration. However, the spectral capability of the detector is sufficient to discriminate the latter, also because count rate of lines at higher energy is strongly decreased because of the decreasing DU quantum efficiency at these energies.

\begin{figure}
\begin{center}
\begin{tabular}{c}
\includegraphics[width=8cm]{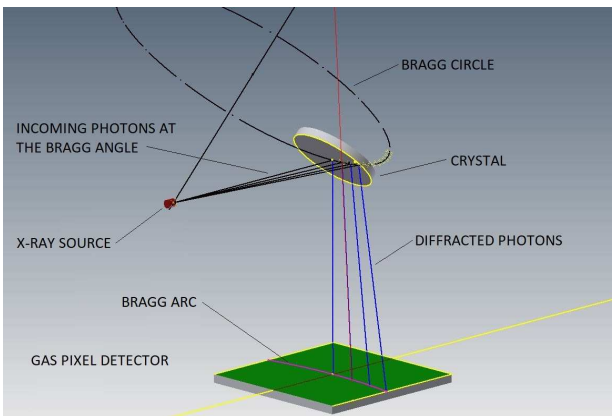}
\end{tabular}
\end{center}
\caption{Geometry of Bragg diffraction for monochromatic photons.}
\label{fig:bragg_strip} 
\end{figure}

\begin{figure}[t]
\begin{center}
\begin{tabular}{c}
\includegraphics[width=6cm]{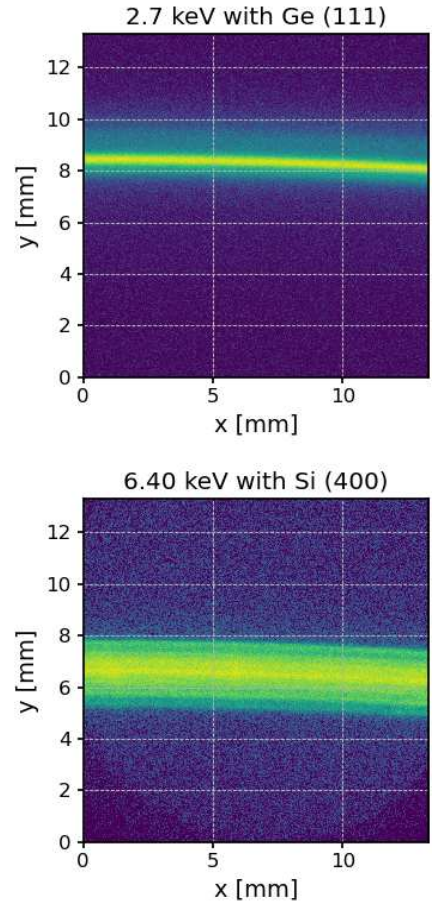}
\end{tabular}
\end{center}
\caption{Image with the ICE test imager of the photons generated with the polarized source without collimators. In this condition, diffracted X-rays forms a ``Bragg'' arc on the detector, whose width depends on both the source and the crystal used. On the top there is the Bragg arc obtained with the X-ray tube with Rh anode and the Ge~(111) crystal; on the bottom there is the Fe X-ray tube with Si~(400) crystal.}
\label{fig:BraggArc}
\end{figure}

\begin{figure*}
\begin{center}
\begin{tabular}{c}
\includegraphics[width=14cm]{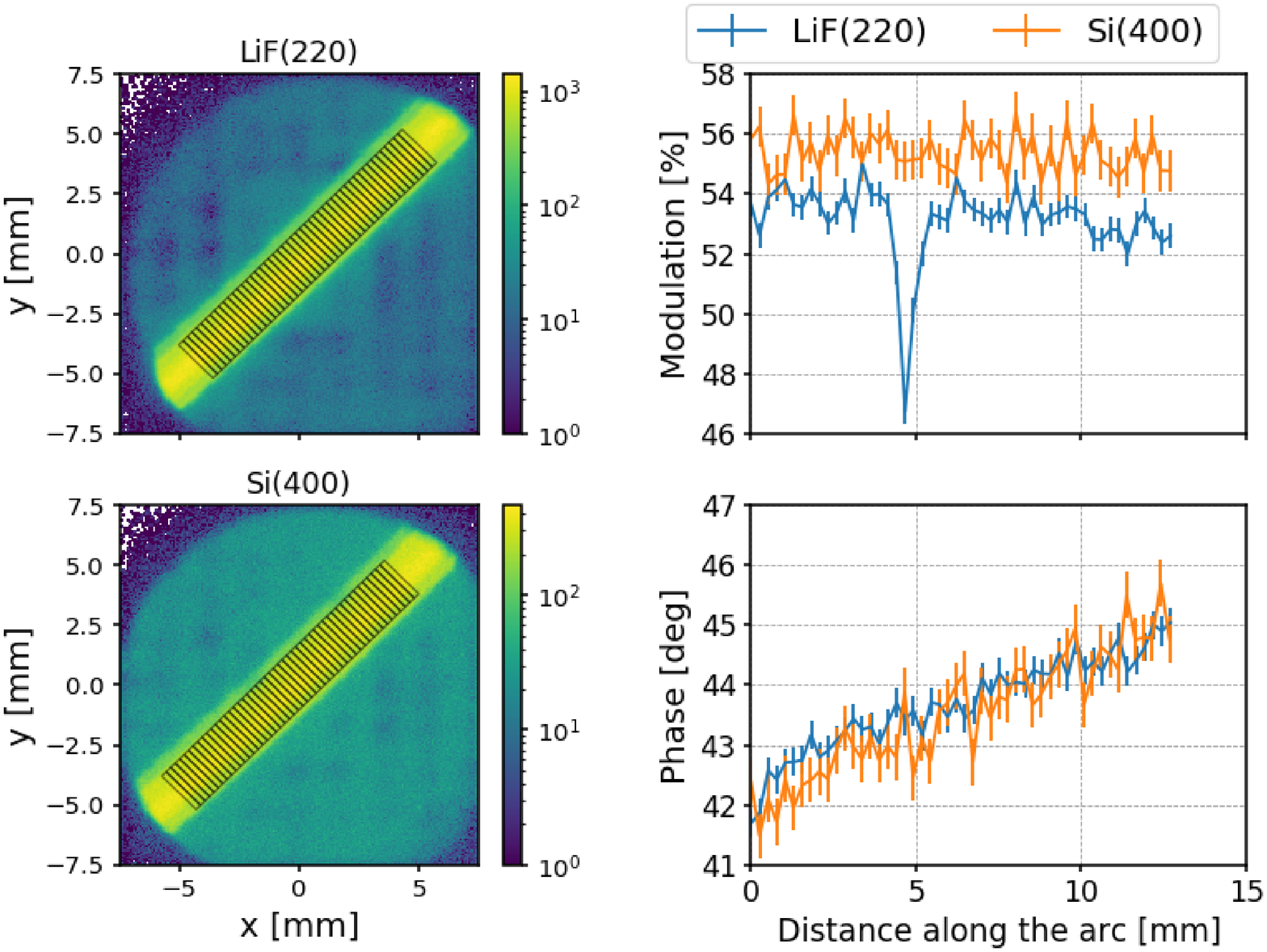}
\end{tabular}
\end{center}
\caption{Study of the polarization along the Bragg arc obtained by diffracting 6.4~keV photons generated with an X-ray tube with Fe anode on LiF~(220) and Si~(400) crystals. The image of the Bragg arc (in the left top and bottom panels) is divided in rectangular regions and for each the modulation and phase measured are reported on the right panels. Modulation of photons diffracted on LiF~(220) presents a drop, whose origin is still under investigation. Modulation of Si (400) is on average higher than that of LiF (220) as the Bragg angle is closer to 45 degrees for the first crystal.}
\label{fig:BraggArc_LiFvsSi} 
\end{figure*}

In addition to the criteria above, crystals were tested also to check the uniformity of polarization of the diffracted X-rays. The incident spectrum can be assumed to be fundamentally monochromatic when X-ray tube fluorescence is tuned with crystal spacing. In this condition, if no collimator is used to constrain incident or diffracted direction, X-rays which satisfy Bragg condition for diffraction produce an arc on the detector (see Figure~\ref{fig:bragg_strip}). Photons along the arc are impinging on the detector with a slightly different incident angle. The width of the arc depends on the X-ray source and the crystal used (see Figure~\ref{fig:BraggArc}). Polarization along the arc is expected to remain constant in degree and  with the polarization angle tangent to the arc. This was verified to be the case for all crystals procured for DU calibration, with the exception of LiF~(220). This is shown in Figure~\ref{fig:BraggArc_LiFvsSi}, where we report the modulation measured with a prototype GPD in spatial bins along the arc when this crystal diffracts 6.4~keV. While the image of the source is uniform and the phase of the modulation remains tangent to the arc as expected, the modulation drops in a region of the arc. For this reason, we used the Si~(400) crystal, which shows a uniform diffraction (see Figure~\ref{fig:BraggArc_LiFvsSi}), instead of the LiF~(220). 

\begin{figure*}
\begin{center}
\begin{tabular}{c}
\includegraphics[width=16cm]{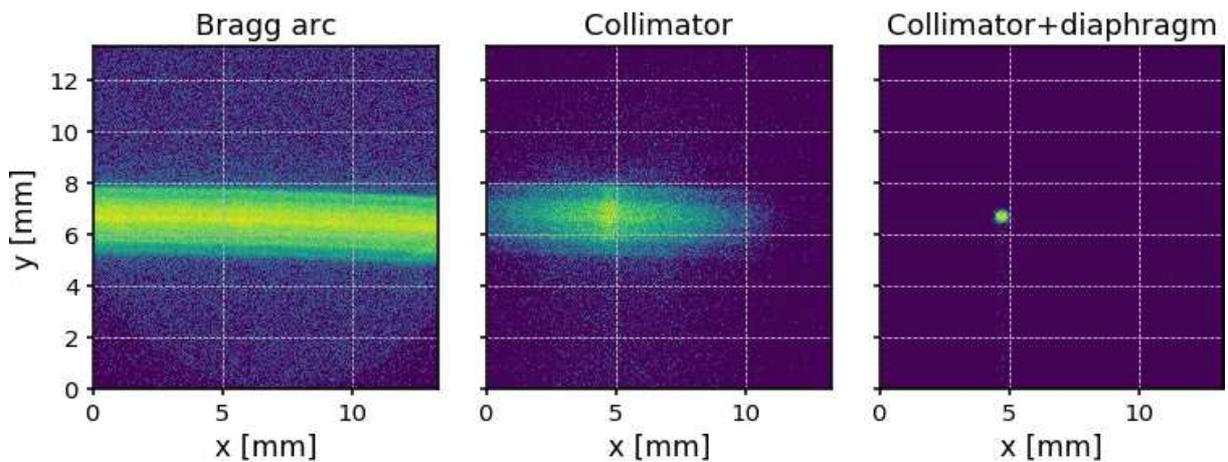} 
\end{tabular}
\end{center}
\caption{Basic steps for polarized source alignment. The Bragg arc is visible when no collimator or diaphragm is present. Adding a collimator, only the Bragg arc corresponding to photon passing through the collimator is visible. The final step is to center a diaphragm in the brightest region of the remaining Bragg arc. At the end of the procedure, the beam is measurable as the line passing through the diaphragm and forming with the crystal plane an angle equal to the Bragg one.}
\label{fig:BraggArc_alignment} 
\end{figure*}

To obtain a beam with a well\--defined and measurable direction and polarization angle, a collimator is used to select a portion of the diffracted and diverging beam, excluding all other directions. This has also the effect of limiting the length of the Bragg arc (see Figure~\ref{fig:BraggArc_alignment}). The final step is to place a diaphragm at the center of the Bragg arc. Such a procedure allows one to derive with metrology the direction of the diffracted beam using the crystal plane, which remains accessible also when the source is assembled, as a reference. In particular, the beam direction lies on the plane orthogonal to the crystal plane, forming with such a plane an angle equal to the Bragg one and passing through the diaphragm (see also Section~\ref{sec:Alignment}). 

\begin{figure}
\begin{center}
\begin{tabular}{c}
\includegraphics[totalheight=5cm]{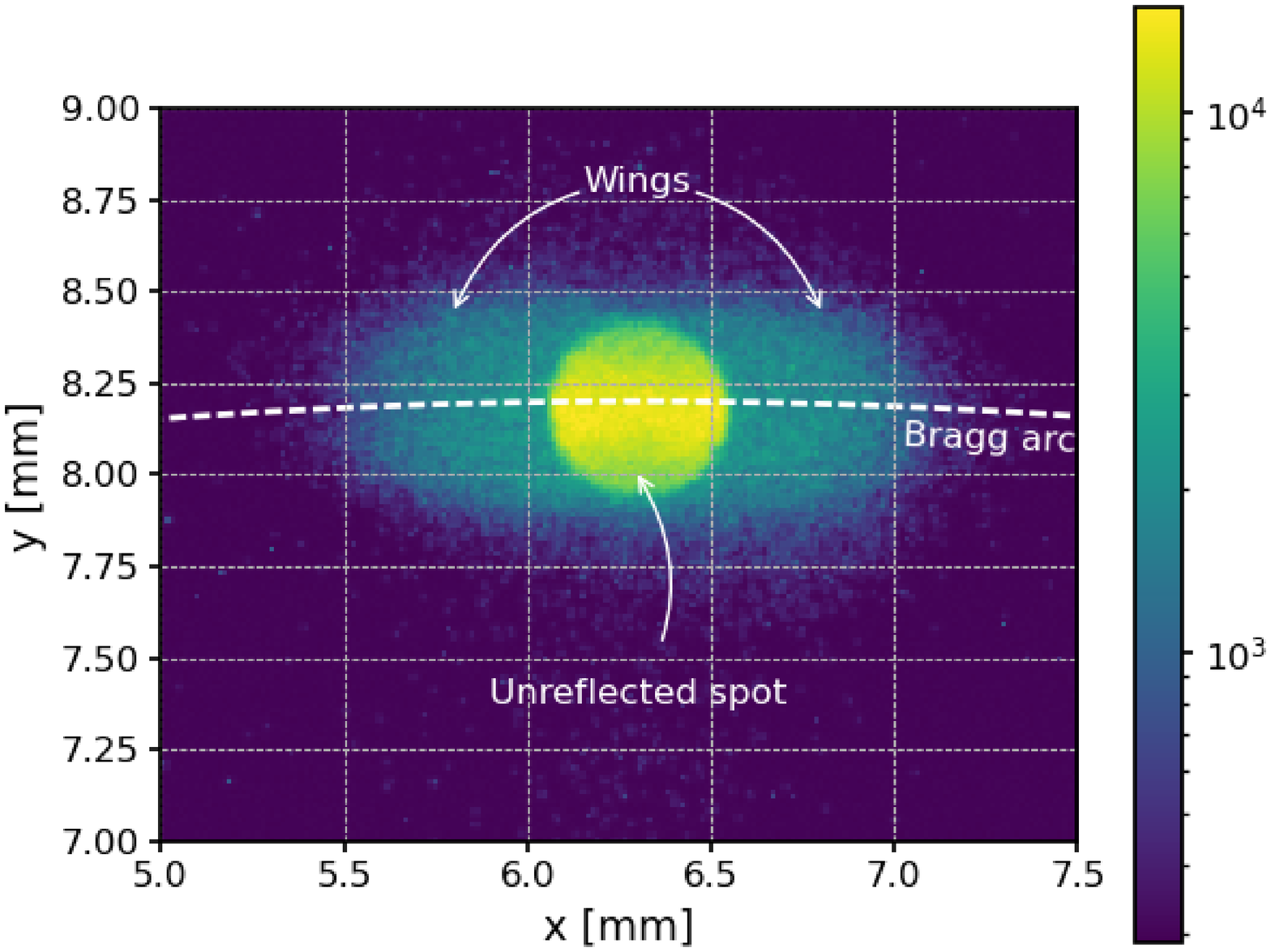}
\\
\includegraphics[totalheight=5cm]{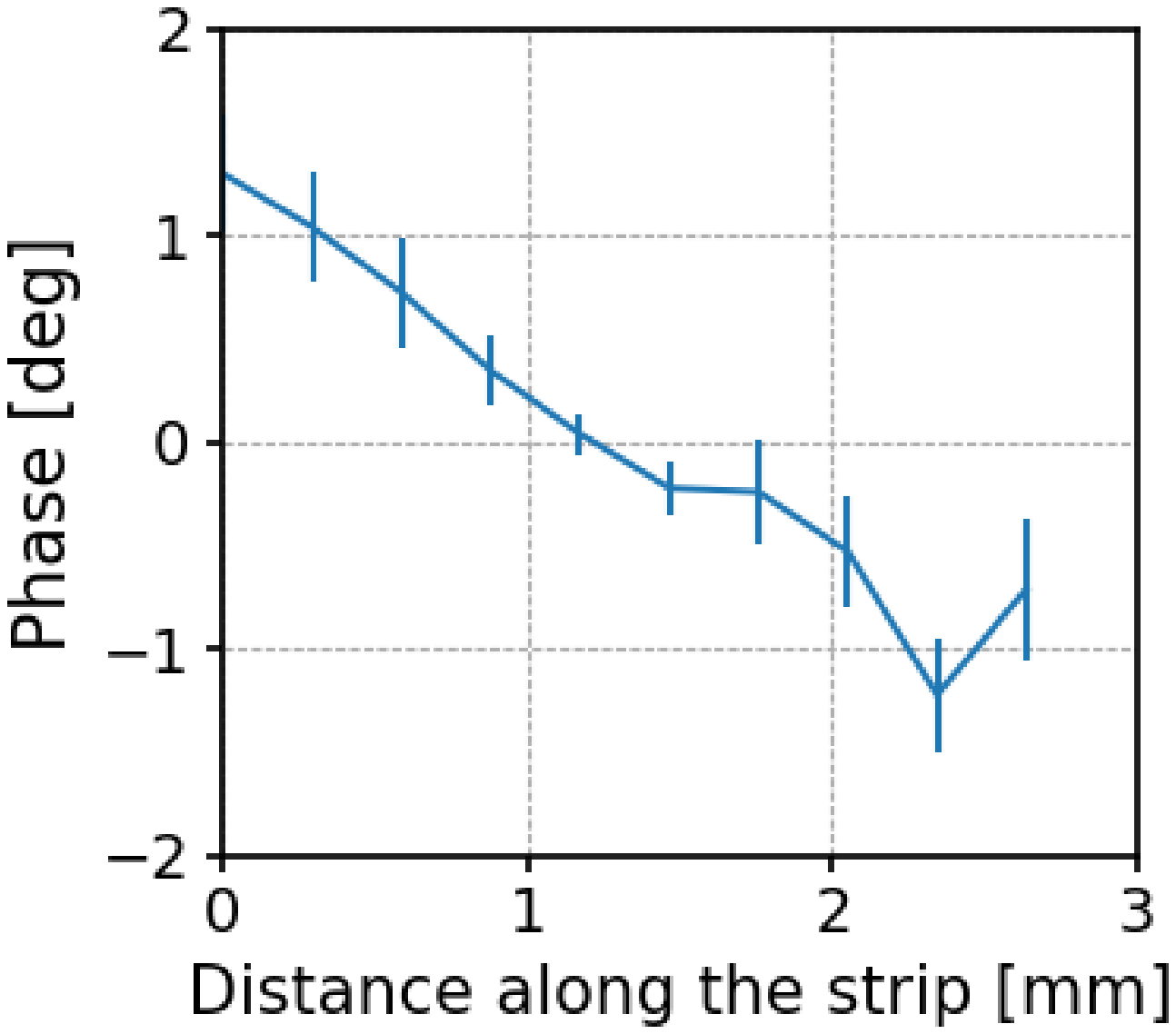}
\end{tabular}
\end{center}
\caption{(Top panel) Image with the ICE test imager of the wings produced by reflection on inner walls of the capillary plates collimator at 2.70~keV. If the image is divided in vertical bins, the angle of polarization is not uniform but it remains tangent to the arc as expected.}
\label{fig:spot_wing} 
\end{figure}

\begin{figure}
\begin{center}
\begin{tabular}{c}
\includegraphics[totalheight=4.5cm]{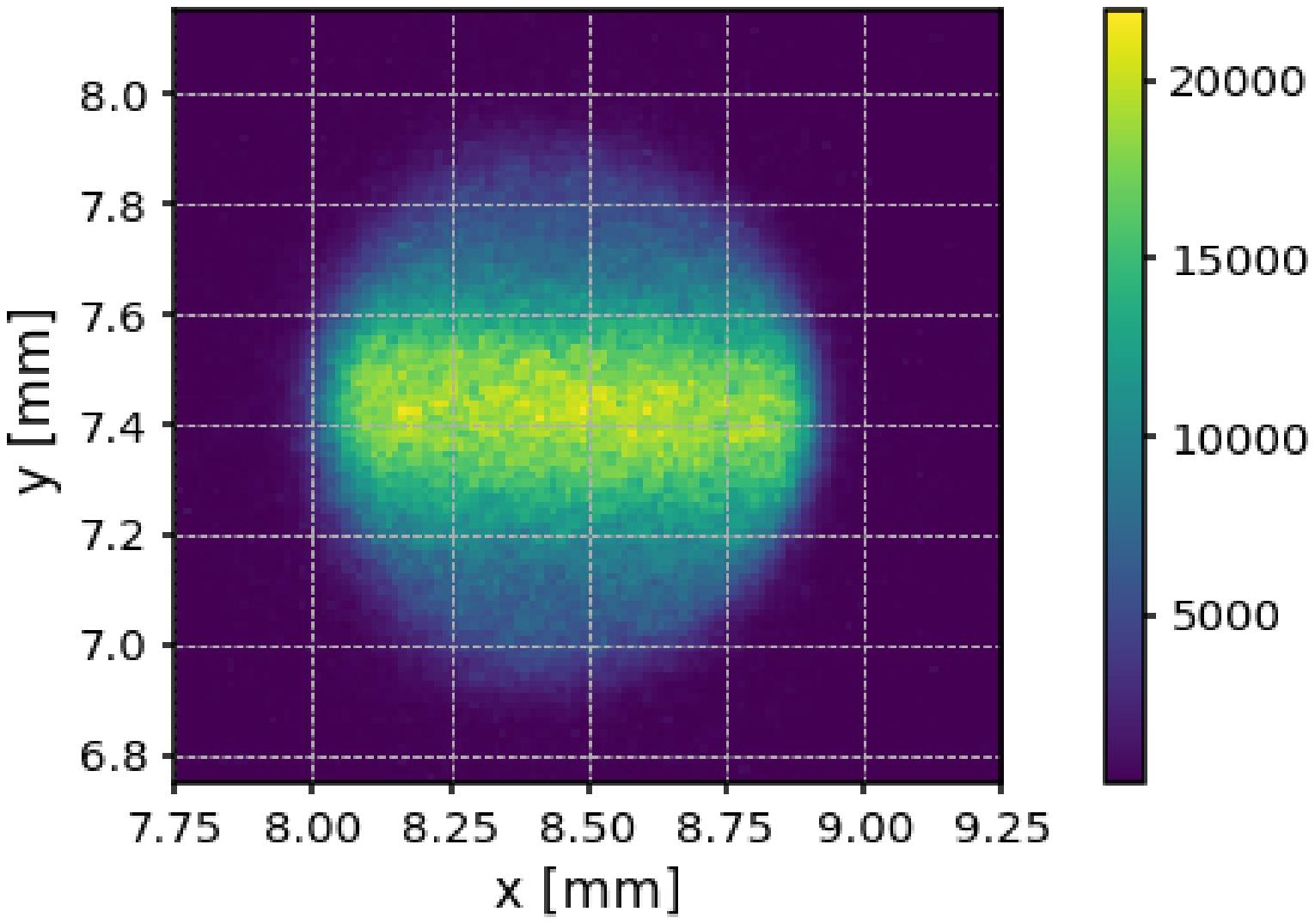}
\\
\includegraphics[totalheight=4.8cm]{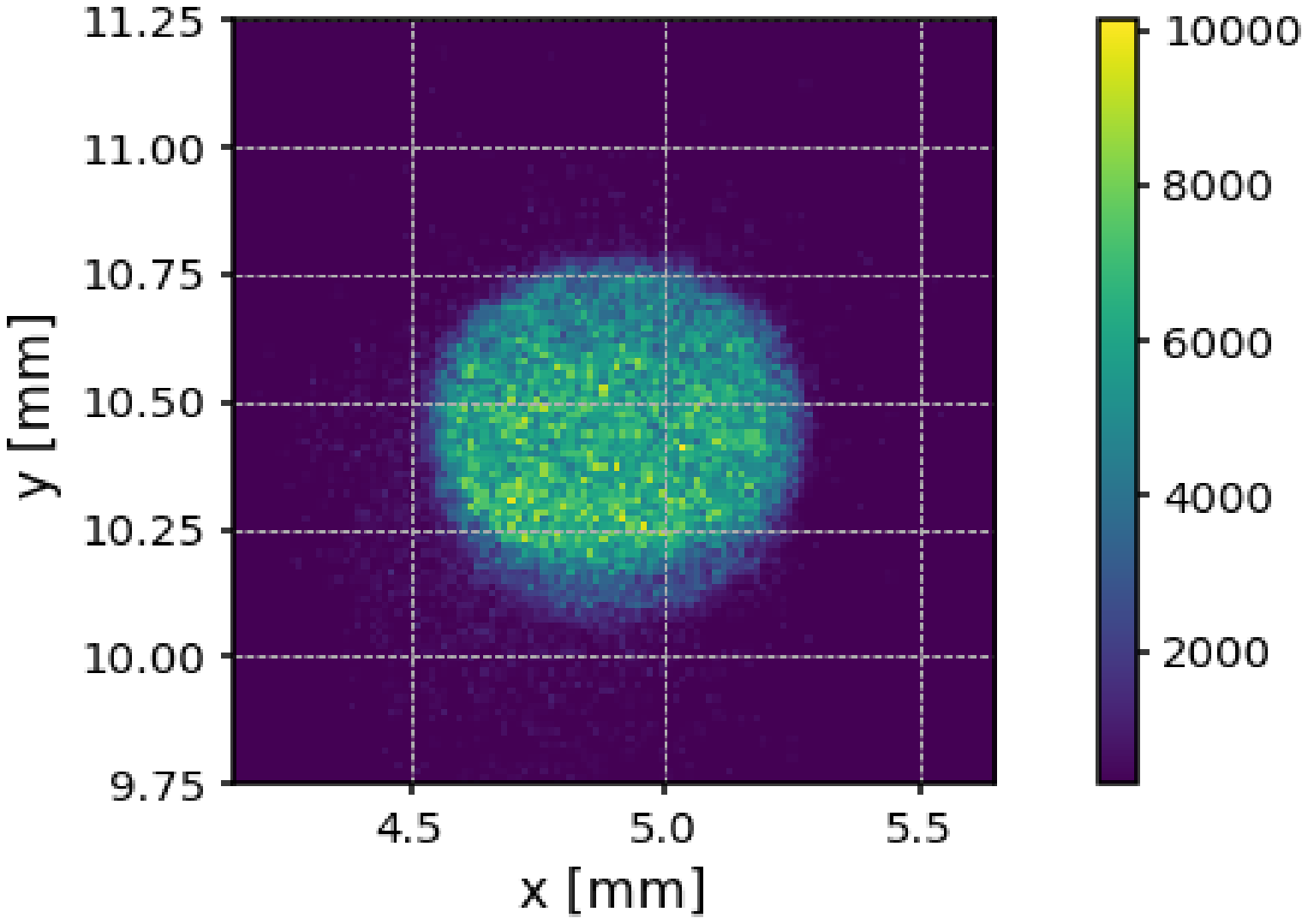}
\end{tabular}
\end{center}
\caption{Spot obtained after the alignment of the polarized source at 2.7 keV (top) and 6.4 keV (bottom) with the T-collimator. Images were acquired with ICE test imager, but the source was moved to the same distance as that used for DU calibration.}
\label{fig:polSpot} 
\end{figure}

In the past, capillary plate collimators were extensively used to constrain the incident or diffracted radiation of this kind of polarized source\cite{Muleri2007,Muleri2008b}. However, the inner walls of such capillary plate collimators reflect a fraction of the diverging diffracted X-rays which would be geometrically stopped by the collimator, especially at lower energy. The final effect is the production of two wings on the side of the main spot in the direction of the Bragg arc, characterized by a small rotation of the polarization angle as expected along the Bragg arc (see Figure~\ref{fig:spot_wing}). Therefore, for DU calibration, we opted to use a mechanical collimator which can completely suppress X-ray reflection (``T-collimator'', see Figure~\ref{fig:PolSource}). Such a collimator is 40 mm long and at its ends has two diaphragms 0.5~mm in diameter, providing a collimation of $\pm$0.7~degrees at zero response. In this case, the spot on the detector is about 0.8~mm in diameter accounting for the distance from the source to the DU (see Figure~\ref{fig:polSpot}). The spot image has memory of the shape of the Bragg arc from which it is extracted: in the case where the Bragg arc is very narrow, as it is at 2.7~keV (see Figure~\ref{fig:BraggArc}), the spot is effectively elliptical. For sources at 2.01 and 3.69~keV, we used the same collimator but with a larger diaphragm of~2 mm to increase the flux from these sources.
 
\begin{figure}
\begin{center}
\begin{tabular}{c}
\includegraphics[width=8cm]{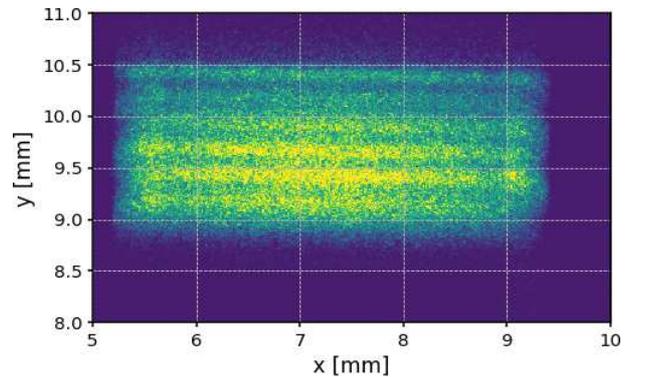}
\end{tabular}
\end{center}
\caption{Example of the scan with the T-collimator. The image is acquired with the ICE test imager. This is equivalent to the middle panel of Figure~\ref{fig:BraggArc_alignment}. Strips are the tracks of the spot during the scan. }
\label{fig:polSourceAlignment} 
\end{figure}

It is worth mentioning that with the use of the T-collimator the procedure described above to center the diaphragm in the brightest part of the Bragg arc has to be slightly modified, as this item works contemporaneously as collimator and diaphragm. The T-collimator is placed in the brightest region of the image obtained by scanning the T-collimator orthogonally to its axis (see Figure~\ref{fig:polSourceAlignment}). Such an image is essentially equivalent to that in the middle panel of Figure~\ref{fig:BraggArc_alignment}.

\section{Procedures}\label{sec:Procedures}

\subsection{Alignment}\label{sec:Alignment}

The detector and the calibration source are aligned before any measurement with a Romer measurement arm (see Figure~\ref{fig:ICE_1}), which measures the relative position and orientation of reference targets to $\sim$10~$\mu$m or $\sim$1~arcmin, respectively. References on the mechanical frame of the GPD are used to derive the position of the detector inside the DU (see Figure~\ref{fig:DU_reference}), and these are compared with references on the sources. Such a procedure is used to align the beam orthogonally to the detector and to preliminarily center the spot on the sensitive area of the detector with an uncertainty of 0.2-1~mm, depending on the source. The position is then refined with X-rays, taking advantage of the imaging capabilities of the detector.

References on the calibration sources depend on its specific design. For direct unpolarized sources, the direction of the beam and its $xy$ position on the detector are derived from the axis of the collimator and its center projected on the GPD plane. The direction of polarization of these sources depends on the geometry of the X-ray tube, and we used the external case of the X-ray tube as a reference. The beam axis of fluorescence unpolarized sources is constrained by the capillary plate collimator. Such an element does not remain accessible when the source is assembled, and then we used the external surface of the source, which is parallel to a high degree ($\approx$0.1~deg) to the collimator, as a reference. Tip/tilt of the DU is adjusted until the beam direction is orthogonal to the GPD plane to better than 0.1 deg.

Alignment of polarized sources is based on the knowledge of the photon diffraction angle and on the measurement of the orientation of the diffracting crystal. The latter is mounted on a holder with a reference plane, which remains accessible after source mounting, and which is co\--planar to better than 0.1~deg with the crystal lattice plane. The beam direction lies on the plane orthogonal to the cystal plane, and it is identified as the line forming the appropriate diffraction angle with the crystal reference plane and passing through the mechanical center of the source diaphragm. The polarization direction is derived as the intersection line between the crystal plane and the detector. The expected polarization angle is then measured as the angle between such a line and the $x$-axis of the GPD as defined from its external references.

\subsection{Dithering}\label{sec:Dithering}

The polarized sources produce a spot of X-rays with diameter $\lesssim$1~mm ($\lesssim$1~mm$^2$), to be compared with a total detector sensitive area of 225~mm$^2$. To calibrate a large area as required by IXPE requirements, the detector was first aligned with the source and then moved continuously to illuminate different regions of the detector. A nearly flat illumination was obtained by moving the DU with the same algorithm that will be used for dithering the IXPE satellite pointing. The movement along $x$ and $y$ at a time $t$ with respect to the centered condition is calculated with
\begin{align*}
 \begin{array}{rcl}
  x&=&a\cos\left(\omega_a t\right)\cos\left(\omega_x t+\frac{\pi}{2}\right)\\
  y&=&a\sin\left(\omega_a t\right)\cos\left(\omega_y t\right) 
 \end{array}
 \qquad \mbox{with} \qquad \omega=\frac{2\pi}{P}
\end{align*}
Periods ($P_a$, $P_x$ and $P_y$) and the amplitude of the movement $a$ were tuned for calibration purposes:
\begin{itemize}
 \item when the measurement is carried out during the night, the length is typically 14~hours and the periods used were the same proposed for the satellite dithering, $P_a=900$~s, $P_x=107$~s, $P_y=127$~s, but radius $a$ was 7~mm instead of about 1.5~mm. A density higher than 40$\times$10$^3$ events/mm$^2$ is typically achieved.
 \item if the measurement is carried out during working hours, the measurement is shorter and then carried out on a smaller (Deep Flat Field) region. The same algorithm is used but periods are 10 times shorter ($P_a=90$~s, $P_x=10.7$~s, $Py=12.7$~s) and radius is 3.25~mm.
\end{itemize}

\begin{figure*}
\begin{center}
\begin{tabular}{c}
\includegraphics[width=14cm]{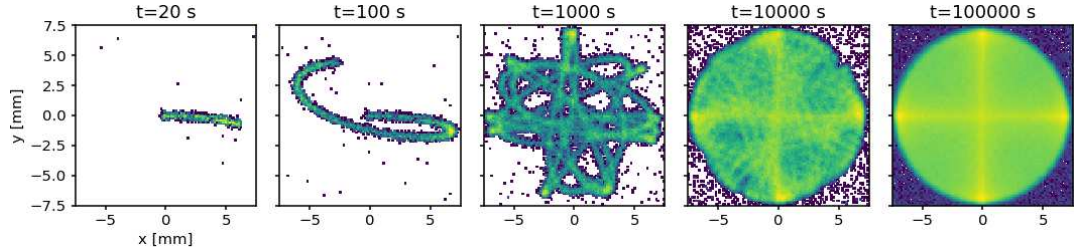}
\end{tabular}
\end{center}
\caption{Example of the movement of the polarized source spot due to dithering in case of a flat field illumination.}
\label{fig:dithering} 
\end{figure*}

\subsection{Air absorption}\label{sec:Absorption}

\begin{figure*}
\begin{center}
\begin{tabular}{c}
\includegraphics[totalheight=5cm]{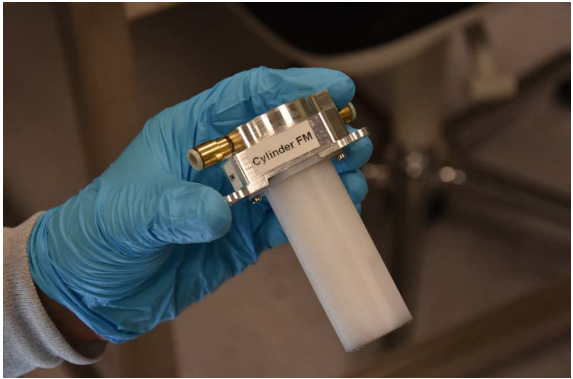}
\includegraphics[totalheight=5cm]{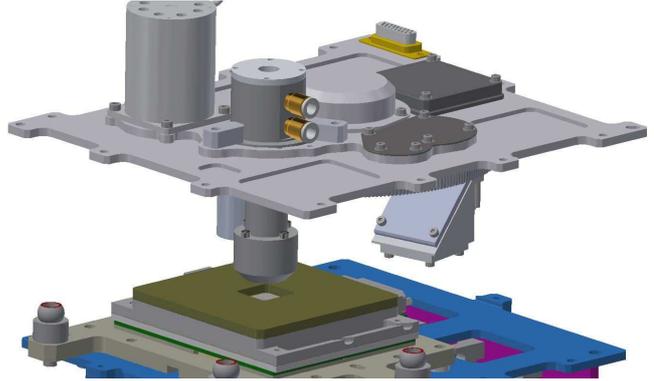}
\end{tabular}
\end{center}
\caption{Cylinder to flow helium in the path from a X-ray source to the detector (left). The figure to the right shows the cylinder mounted inside the DU during calibration.}
\label{fig:HeCylinder} 
\end{figure*}

X-rays in the IXPE energy range are strongly absorbed by air and this causes both a reduction of source flux and a change in the source spectrum, since absorption is strongly energy-dependent. In our set-up, absorption occurs along the path of the X-rays from the window of the tube to the window of the detector. To reduce it, helium is flowed inside the volume of the mechanical assembly mounted on the X-ray tube and the source is moved as close as possible to the DU before starting the measurement. The DU stray-light collimator, which extends from the DU for $\sim$260~mm to prevent X-rays outside the IXPE telescope to impinge on the detector, is removed at the beginning of DU calibration. Nevertheless, the air path from the unit top lid to the detector is sufficient to absorb more than 50\% of the photons at energies below 4~keV and more than 90\% below~2.8 keV. To avoid such a large deficiency, a cylinder is inserted inside the DU to saturate with helium also the path to the detector which is internal to the DU. The cylinder is sealed with two windows of 4~$\mu$m-thick polypropylene film which are almost transparent to X-rays in the IXPE energy band. A picture of the cylinder and its mounting inside the DU are shown in Figure~\ref{fig:HeCylinder}.  

While the primary function of the helium cylinder was to avoid air absorption, different versions were used also to constrain the source spot. This varied significantly for the different measurements. For example, in calibration of the response to unpolarized X-rays or for Instrument testing, sources produce essentially isotropic beams (see Section~\ref{sec:Unpolarized}), and then no diaphragm was used for Flat Fields, whereas one with 3.3~mm radius was used for the DFF. In the latter case, the diaphragm was mounted at the end of the helium cylinder to place it as close as possible to the detector and increase the sharpness of the spot notwithstanding the finite spot size of the X-ray source.

\subsection{Contamination}

The requirements on the maximum allowable particulate and molecular contamination (PAC and MOC) on the DU optical elements, that is, essentially the X-ray entrance window of the GPD, are 2000~ppm and 4$\times$10$^{-6}$~g/cm$^2$, respectively. Contamination during calibrations is avoided by covering the X-ray aperture of the DU with a filter made of 4~$\mu$m-thick polypropylene, which is essentially transparent in the energy range of interest for calibration. The filter is removed only for particular measurements, like the DU absolute efficiency, and during this time witness samples are exposed for monitoring the contamination. The analysis of such samples and the periodic use of a particle counter confrimed the achievement of the contamination control requirements.

\section{Conclusion}

The Instrument Calibration Equipment (or ICE) and the AIV-T Calibration Equipment (ACE) were specifically built for the calibration and testing of the unique polarization-sensitive focal plane detectors of the Imaging X-ray Polarimetry Explorer mission. These detectors demanded extensive characterization which was carried out at INAF-IAPS in Rome, Italy, for more than 6~months. The ICE features custom sources built for the peculiar needs of IXPE and which are able to generate polarized and unpolarized X-ray photons in the energy range between $\lesssim$2~keV and $\gtrsim$8~keV. Depending on the source, the energy of the radiation can be truly monochromatic, and there is a large choice of spot size and source flux. Motorized and manual stages allow for the alignment of source components to increase the flux and constrain the direction of the beam. Moreover, these stages are used to align the source and the detector with a measurement arm and to move the beam over the detector sensitive area if needed. Test detectors, a CCD imager and a SDD spectrometer, are available to characterize the beam before the calibration. 

The ACE was designed to contemporaneously illuminate with X-ray sources the three detectors which comprise the focal plane of the IXPE observatory. For this purpose, the ACE can host up to three of the calibration sources built for the ICE, with a subset of the stages available on the ICE for the alignment and movement of the beam. In fact, this allowed us to carry out some of the most time-consuming calibrations on both the ICE and the ACE contemporaneously, which was convenient in the tight IXPE schedule. 
 
Contrary to larger calibration facilities, the ICE and ACE are operated in air. This choice is driven by the need to simplify the operations and increases the flexibility of the equipment. Measurement at low energy are possible flowing helium along the photon path, which makes negligible the absorption of X-rays in the working energy range. The overall system is extremely reliable: taking advantage of the detector capability of switching off safely in case of power outage, IXPE calibrations were carried out 24 hour per day, 7 days per week continuously for several months. All of these characteristics make the ICE and the ACE ideal facilities for the calibration of X-ray detectors, especially if this requires custom operations, and indeed they are under evaluation for other projects.

\subsection*{Acknowledgments}

The Italian contribution to the IXPE mission is supported by the Italian Space Agency (ASI) through the contract ASI-OHBI-2017-12-I.0, the agreements ASI-INAF-2017-12-H0 and ASI-INFN-2017.13-H0, and its Space Science Data Center (SSDC), and by the Istituto Nazionale di Astrofisica (INAF) and the Istituto Nazionale di Fisica Nucleare (INFN) in Italy. The USA contribution to IXPE is supported by NASA under the Small Explorer Program.

\bibliography{References.bib}
\bibliographystyle{elsarticle-num}

\end{document}